\documentclass[12pt,preprint]{aastex}

\usepackage{graphicx,epstopdf}

\usepackage{color}

\def\BB{{\bf {B}}}

\def\EE{{\bf {E}}}

\def\xx{{\bf {x}}}
\def\ww{{\bf {w}}}

\shorttitle{3D null point reconnection in the corona}
\shortauthors{Pontin {\it et al.}}

\begin{document}

\title{On the nature of reconnection at a solar coronal null point above a separatrix dome}


\author{D.~I.~Pontin}
\affil{Division of Mathematics, University of Dundee, UK}
\email{dpontin@maths.dundee.ac.uk}

\author{E.~R.~Priest}
\affil{School of Mathematics and Statistics, University of St Andrews, UK}

\author{K.~Galsgaard}
\affil{Niels Bohr Institute, Copenhagen, Denmark}

\begin{abstract}
Three-dimensional magnetic null points are ubiquitous in the solar corona, and in any generic mixed-polarity magnetic field. We consider magnetic reconnection at an isolated coronal null point, whose fan field lines form a dome structure. We demonstrate using analytical and computational models several features of spine-fan reconnection at such a null, including the fact that  substantial magnetic  flux transfer from one region of field line connectivity to another can occur. The flux transfer occurs across the current sheet that forms around the null point during spine-fan reconnection, and there is no separator present. 
{Also, flipping of magnetic field lines takes place in a manner similar to that observed in quasi-separatrix layer or slip-running reconnection.}
\end{abstract}



\section{Introduction}
\begin{figure}[t]
\centering
\includegraphics[width=3.5in]{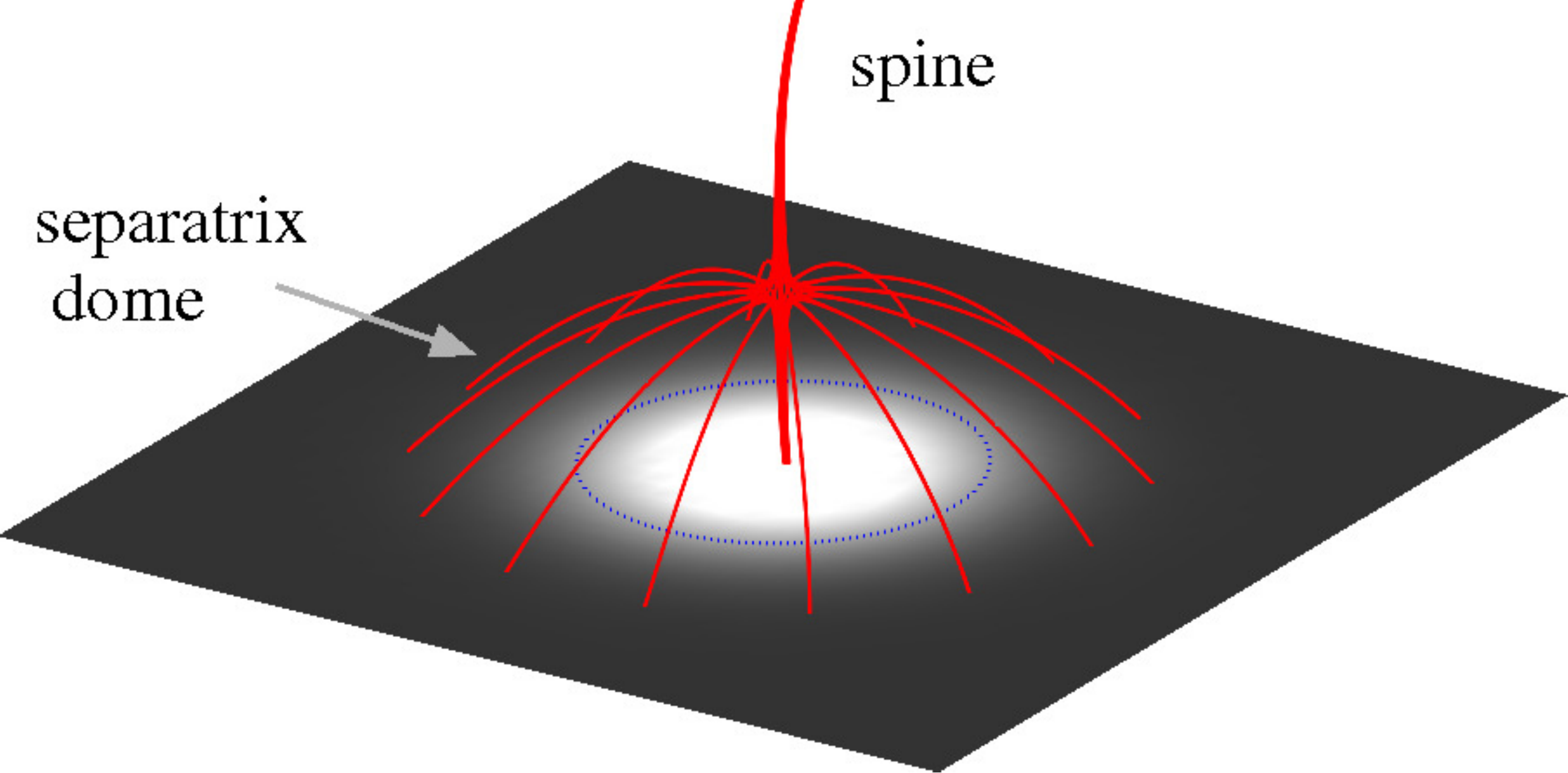}
\caption{Magnetic field lines outlining the spine and fan structures associated with a magnetic null, located in a separatrix dome above a parasitic polarity. Shading on the lower surface represents the normal component of ${\bf B}$ and the dashed line marks the polarity inversion line.}
\label{domefig}
\end{figure}

It is known that magnetic reconnection occurs frequently in astrophysical plasmas even though such plasmas typically have extremely low dissipation. The resolution of this apparent contradiction is that enormous currents must form on very small length scales. The conditions under which thin current sheets form -- and reconnection subsequently occurs -- remains a hotly debated topic. This debate has been fuelled by the recent discovery that solar coronal magnetic fields have a highly complex topological structure, down to scales much smaller than previously thought. One generic feature of such complex fields  found in astrophysical plasmas is the presence of magnetic null points (points in space at which the magnetic field vanishes). 
The structure of the magnetic field in the vicinity of a null point is characterised by a \emph{spine line} along which field lines approach (or recede from) the null, and a \emph{fan surface} along which field lines recede from (or approach) the null -- see Figure \ref{domefig}. The fan is a \emph{separatrix surface} that divides the local volume into two topologically distinct regions with respect to the magnetic field line connectivity. 
Magnetic null points and their associated separatrix surfaces have long been proposed as locations for the formation of thin current layers, and the conversion of magnetic energy via reconnection  \cite[e.g.][]{lau1990, priest1996,antiochos1996, priest2002}. 

The traditional picture of magnetic reconnection involves the transport of magnetic flux across the separatrices associated with a 2D magnetic $X$-point. As discussed in Section \ref{recsec} below, the complexity and diversity of the reconnection process is greatly increased when one moves to three dimensions -- clearly the relevant case for astrophysical plasmas. One mechanism for 3D reconnection involves the formation of a current layer along a separator field line (a magnetic field line joining a pair of 3D nulls), with magnetic flux being transported across the pair of separatrix surfaces associated with the nulls. It might seem natural to suggest that meaningful flux transfer only occurs at separators.  Of course flux transfer in 3D occurs only across a surface and so cannot occur at a single  point (such as a null point). However, substantial flux transfer can occur across the current sheet that forms around a null point when 3D spine-fan reconnection (see Section \ref{backsec}) occurs.

There is observational evidence to support the importance of nulls for energetic processes in space plasmas. Recent studies show that the magnetic carpet field of the quiet Sun must consist of a complex array of magnetic nulls points, separatrix surfaces, and separators \citep{schrijver2002,regnier2008,longcope2009}. Indeed, it has been suggested that reconnection at such nulls plays an important role in solar flares, CMEs, jets and bright points \citep[e.g.][]{barnes2007,masson2009,pariat2009,zhang2012,morenoinsertis2013}. There are also observations of flare ribbons that appear to be associated with particle acceleration during null point reconnection \citep{zuccarello2009,liu2011}. In addition, recent observations in the Earth's magnetosphere show that there are many more nulls present than initially expected, both in the cusp regions \citep[e.g.][]{dorelli2007} and in the turbulent plasma of the magnetotail \citep[e.g.][]{xiao2007}. 

In this study we focus on reconnection at magnetic nulls in the solar corona. That is, we consider nulls in a half-space above a perfectly conducting photospheric plane. The existence of null points in the corona is dependent on the magnetic field distribution on the photosphere (as well as dynamical processes occurring in the corona if the field is not in equilibrium).
The simplest generic configuration in which a null may appear in the volume is shown in Figure \ref{domefig}, and  involves an isolated null whose fan field lines form a separatrix dome that connects down to the photosphere (and which therefore do not connect to any other null to form a separator).
The separatrix surface divides flux that is locally closed from flux that is locally open (but may close at some other distant location on the photosphere). Such a dome structure exists above any parasitic polarity region in which a region of one sign of vertical magnetic flux is surrounded by a region of opposite signed -- and greater total -- flux. This characteristic separatrix dome configuration is the structure most frequently observed in the coronal field extrapolations discussed above. It is also a characteristic structure often associated with X-ray jets and solar flares.

In this paper we discuss  how the magnetic field of the corona is restructured by reconnection at a 3D coronal null. 
The paper is arranged as follows. In Section \ref{backsec} we review previous results on current sheet formation at null points and 3D reconnection. In Section \ref{toysec} we introduce a simple model that demonstrates the effects of reconnection at a coronal null on the coronal field structure. In Section \ref{numsec} we show that the results carry forward when a full MHD evolution is followed numerically, and we present a discussion in Section \ref{concsec}.

\section{Background: Current Sheets and Reconnection at 3D Nulls}\label{backsec}
\subsection{Formation of Current Sheets at 3D Nulls}\label{jsheetsec}
Magnetic reconnection always occurs in strong current concentrations, usually in thin current sheets. 
Such current sheets may form dynamically at nulls points, separators or quasi-separatrix layers, in response to a driving of the system or to a spontaneous relaxation process. There also exist additional mechanisms of current sheet formation, including ideal MHD instabilities. Here we focus on the occurrence of current sheets at 3D nulls. 

There are a number of compelling arguments as to why current sheets should naturally form at 3D nulls. First,  various studies have demonstrated the build-up of currents close to nulls. For example, linearising the magnetic field (${\bf B}$) and plasma flow (${\bf v}$) about the null in an open system it has been shown that gradients of ${\bf B}$ and ${\bf v}$ tend to become singular \citep{bulanov1984,bulanov1997,klapper1996,parnell1997}. 
Furthermore, \cite{hornig1996} proved that  certain evolutions of the magnetic field in the vicinity of a 3D null are prohibited by an ideal evolution. In particular, the ratio between any pair of eigenvalues of the Jacobian matrix of ${\bf B}$ evaluated at the null must remain fixed. It can be demonstrated that a variation in these ratios is a natural consequence of external perturbations focussing around the null \citep{pontinbhat2007a}. This strongly suggests that, in the absence of dissipation, a singular current sheet will form at the null, and that in the presence of a small but finite dissipation a thin, intense current layer will form. The formation of singular current layers at nulls in a line-tied volume during an ideal relaxation has been demonstrated numerically 
\citep{pontincraig2005,pontin2012b}.
Furthermore, the formation of thin (but finite) current sheets in a resistive plasma has been demonstrated both in numerical simulations \citep{pontinbhat2007a} and in laboratory experiments \citep{bogdanov1994, frank2001}.

In each of the studies discussed above, the current sheet that forms at the null involves a local collapse of the spine and fan towards one another. The current sheet forms at an angle intermediate between the global orientations of the spine and fan and the current vector is oriented orthogonal to the spine line (see e.g. Figure 6 of \cite{pontin2012a}). Current layers focussed on either the spine or fan can also form in response to rotational perturbations centred on the spine \citep{galsgaardpriest2003,pontingalsgaard2007}. There is an associated spiralling of field lines in the current layers, within which torsional spine or torsional fan reconnection (see below) take place. These reconnection modes are not our focus here. Indeed, there are also indications that in the absence of resistivity a singular current may form with the current vector parallel to the spine line, associated with an increasingly tight spiralling of field lines around the spine \citep{fuentes2012}.

\subsection{Properties of magnetic reconnection in three dimensions	}\label{recsec}
{Fundamentally, magnetic reconnection involves a breakdown in the magnetic connection between plasma elements. In three dimensions this occurs in general when a spatially-localised  component of the electric field parallel to the magnetic field ($E_\|$) exists \citep{schindler1988,hesse1988}. The rate of reconnection is defined as the maximal value of
\begin{equation}\label{recratedef}
\Phi=\int E_\| ds,
\end{equation}
over all magnetic field lines threading the non-ideal region (region within which $E_\| \neq 0$), where $s$ is a parameter along the field lines.
In general, the equation
\begin{equation}\label{idealev}
\frac{\partial {\bf B}}{\partial t}- \nabla\times (\ww \times \BB)={\bf 0},
\end{equation}
describes the ideal evolution of a magnetic field $\BB$, where ${\bf w}$ is a flux transport velocity. If, for a given magnetic field evolution, a smooth flow ${\bf w}$ exists then  the topology of the magnetic field is preserved, and no reconnection occurs (the magnetic flux is {\it frozen into} the flow $\ww$). The properties of 3D reconnection are then crucially different from the simplified 2D picture in the following ways. }

{In 2D, a flux transport velocity $\ww$ exists everywhere except at an $X$-point or $O$-point, where it is singular if ${\bf E}\neq{\bf 0}$ there. The singularity of $\ww$ at an $X$-point is a signature of the fact that each magnetic field line is cut and rejoined in the instant that it passes through the $X$-point (and separatrix). Since $\ww$ is smooth and continuous everywhere else, field lines evolve as if they are reconnected  \emph{only} at the $X$-point,  in a one-to-one pairwise fashion.}

{In 3D,  reconnection occurs within a spatially localised region of $E_\|$, and one can prove that no flux transport velocity $\ww$ exists for the flux threading this non-ideal region \citep[see][]{priest2003a}.
As a result, within any infinitesimal time interval $\delta t$, {\it every} field line threading the non-ideal region changes its connectivity. Therefore, magnetic field lines traced from footpoints comoving in the ideal flow appear to split {\it as soon as they enter the non-ideal region}, and their connectivity changes {\it continually and continuously} as they pass through the non-ideal region.}
This means that the restructuring of the magnetic field no longer involves magnetic field lines being reconnected in a one-to-one fashion as in the 2D case. We emphasise that the above holds true for \emph{all} reconnection processes in 3D, independent of the local field structure.

While no single flux velocity can describe the motion of magnetic field lines during 3D reconnection, one can nevertheless replace the concept by a flux velocity pair $\ww_{in}$ and $\ww_{out}$ \citep{priest2003a}. This can be a useful tool to visualise the evolution of field lines during the reconnection process.  $\ww_{in}$ represents the evolution of field lines entering the non-ideal region (with respect to the direction of $\BB$) traced from footpoints co-moving in the ideal flow. Correspondingly, $\ww_{out}$ represents the velocity of field lines exiting the non-ideal region, traced from ideal co-moving footpoints.
The continuous change of field line connectivity means that field lines that are traced through and beyond the non-ideal region move at some virtual flow velocity that is not related to the local plasma velocity.  This often leads to an apparent `flipping' of magnetic field lines (as first suggested by \cite{priest1992}). This velocity has been ascribed physical significance in the idea of `slip-running reconnection' \citep{aulanier2006}.

\subsection{Modes of reconnection at 3D null points}\label{nullrecsec}
There are different characteristic modes through which reconnection can take place in 3D. Here we focus on the modes of reconnection at 3D nulls, which were categorised by \cite{priest2009}. 
The relevant reconnection mode that we consider here is {\it spine-fan reconnection}. This occurs in a current sheet localised around the null that forms in response to a collapse process as discussed in Section \ref{jsheetsec}. Such a collapse is initiated by any shear disturbance of the spine or fan (i.e.~any disturbance locally orthogonal to the spine/fan field lines). The result is a transfer of magnetic flux across the fan separatrix surface as well as past the spine line. Tracing field lines from co-moving ideal footpoints transported past the spine, one observes a flipping of these field lines around the fan plane. So, while flipping of field lines is a natural signature of QSL reconnection, one should note that it also occurs in 3D spine-fan reconnection at a null point. This is a direct result of the fundamental theorem regarding the non-existence of a flux transporting flow as discussed above.

If the plasma is incompressible, pure spine or fan reconnection modes may exist \citep{priest1996,craig1996} in which the current extends along the spine or fan, respectively. {However, so far only the incompressible fan reconnection solutions have been shown to be dynamically accessible, and when the incompressibility assumption is relaxed it turns out that the plasma pressure gradient is too weak to oppose the collapse of the null and formation of a fully localised current sheet at the null \cite[see section IV\,B of][]{pontinbhat2007b}.}


{Two additional reconnection modes result from rotational motions centred on the spine. These lead to current concentrations localised to either the spine or the fan, within which a type of rotational slippage known, respectively, as either {\it torsional fan reconnection} or {\it torsional spine reconnection} occurs \citep{galsgaardpriest2003,pontin2011c,wyper2010}.}
In neither of these reconnection modes is there any flux transfer across the separatrix surface, and so they are not relevant to the present study.

\section{Simple model}\label{toysec}
\subsection{Model setup}
\begin{figure}[t]
\centering
\includegraphics[width=4in]{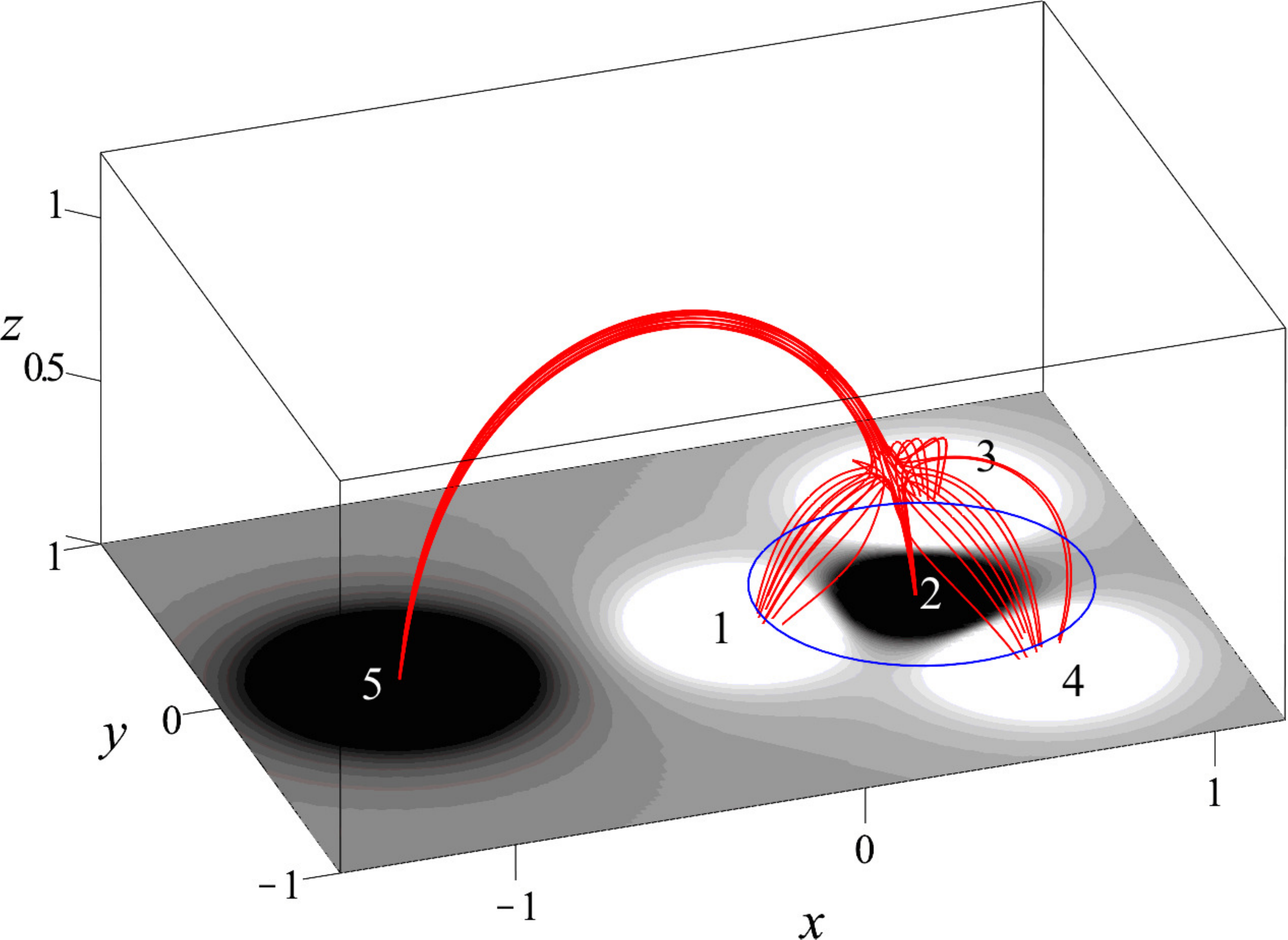}
\caption{Field lines of the simple model magnetic field (described in Section \ref{toysec}) outlining the spine and fan of the magnetic null, at $t=0$. The shading on the lower surface represents the normal component of the magnetic field there. The circle on the lower surface shows the approximate location of the footprint of the separatrix dome (i.e., fan separatrix surface). The numbers refer to source numbering discussed in the text.}
\label{toy_blines}
\end{figure}

In this section we present a simple model that demonstrates the fundamental properties of magnetic reconnection as it occurs at a 3D null point in the solar corona. We begin with a potential magnetic field due to 5 photospheric flux patches, constructed by placing five magnetic point charges at locations outwith our domain of interest. Specifically, we restrict our studies to the half-space $z>0$, where $z=0$ represents the photosphere, and place all point charges at $z<0$. The resultant magnetic field is given by 
\begin{equation}\label{b_pot}
\BB_p=\sum_{i=1}^n \epsilon_i \frac{\xx-\xx_i}{|\xx-\xx_i|^3}
\end{equation}
where $n=5$ and where $\xx_i$ are the locations and $\epsilon_i$ are the strengths of the point charges. Here we take $\{\epsilon_1,\epsilon_2,\epsilon_3,\epsilon_4,\epsilon_5\}=\{0.01,-0.01,0.01,0.01,-0.02\}$ and $\xx_1=(0,0,-0.1),~\xx_2=(0.5,0,-0.1),~\xx_3=(0.7,0.5,-0.1),~\xx_4=(0.7,-0.45,-0.1),~\xx_5=(-1,0,-0.1)$. We refer hereafter to the flux patch in the photosphere ($z=0$) associated with the charge located at $\xx_i$ as `source $i$'. Since the negative source 2 is surrounded by three positive sources, a null point is naturally located above source 2. The null is located at $\xx=\xx_N=(0.416,0.00478,0.390)$. The spine of the null intersects the photosphere within sources 2 and 5, while the footprint of the dome-shaped fan surface links sources 1, 3 and 4 (see Figure \ref{toy_blines}).

\begin{figure}[t]
\centering
(a)\includegraphics[width=3.in]{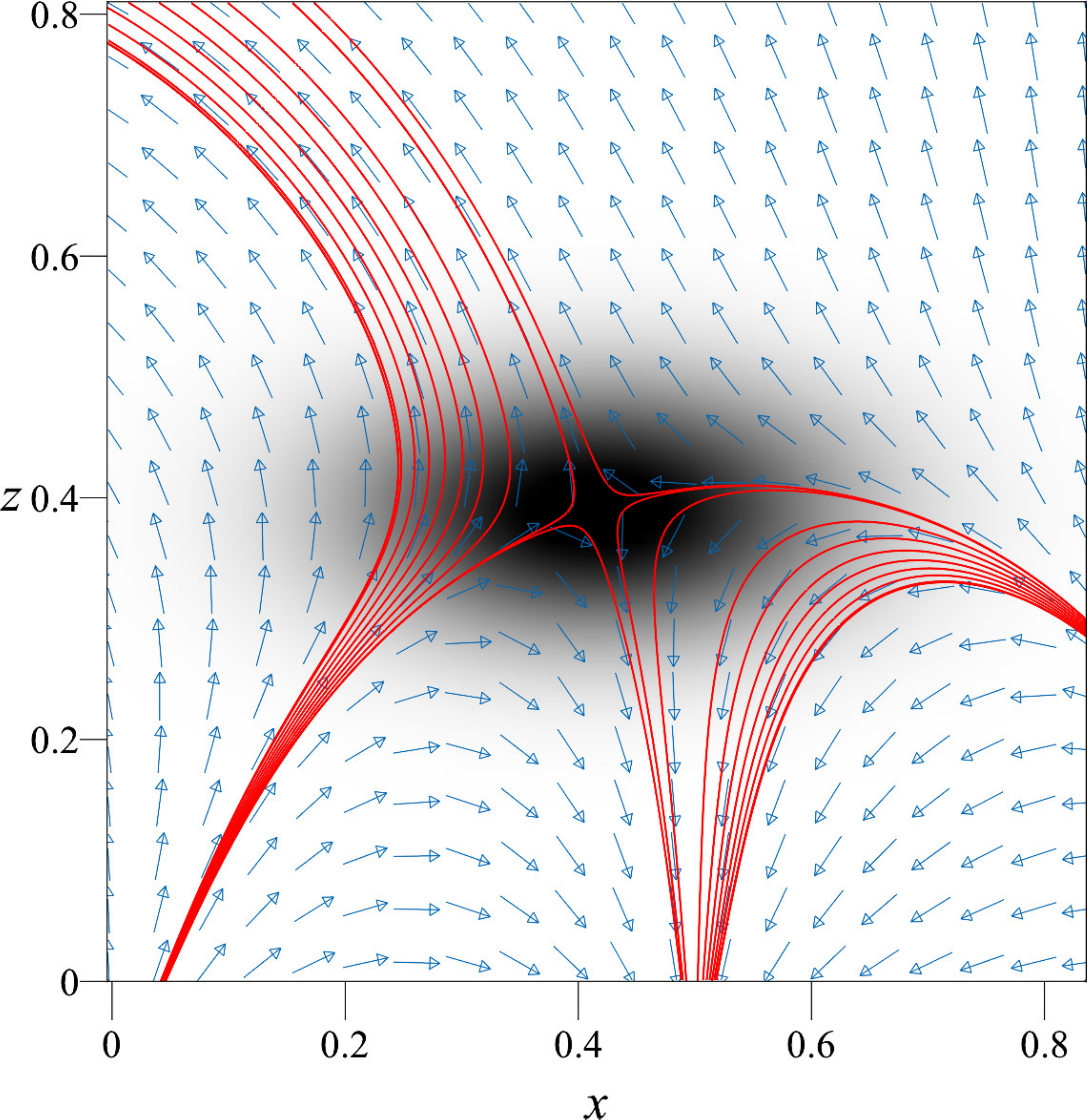}
(b)\includegraphics[width=3.in]{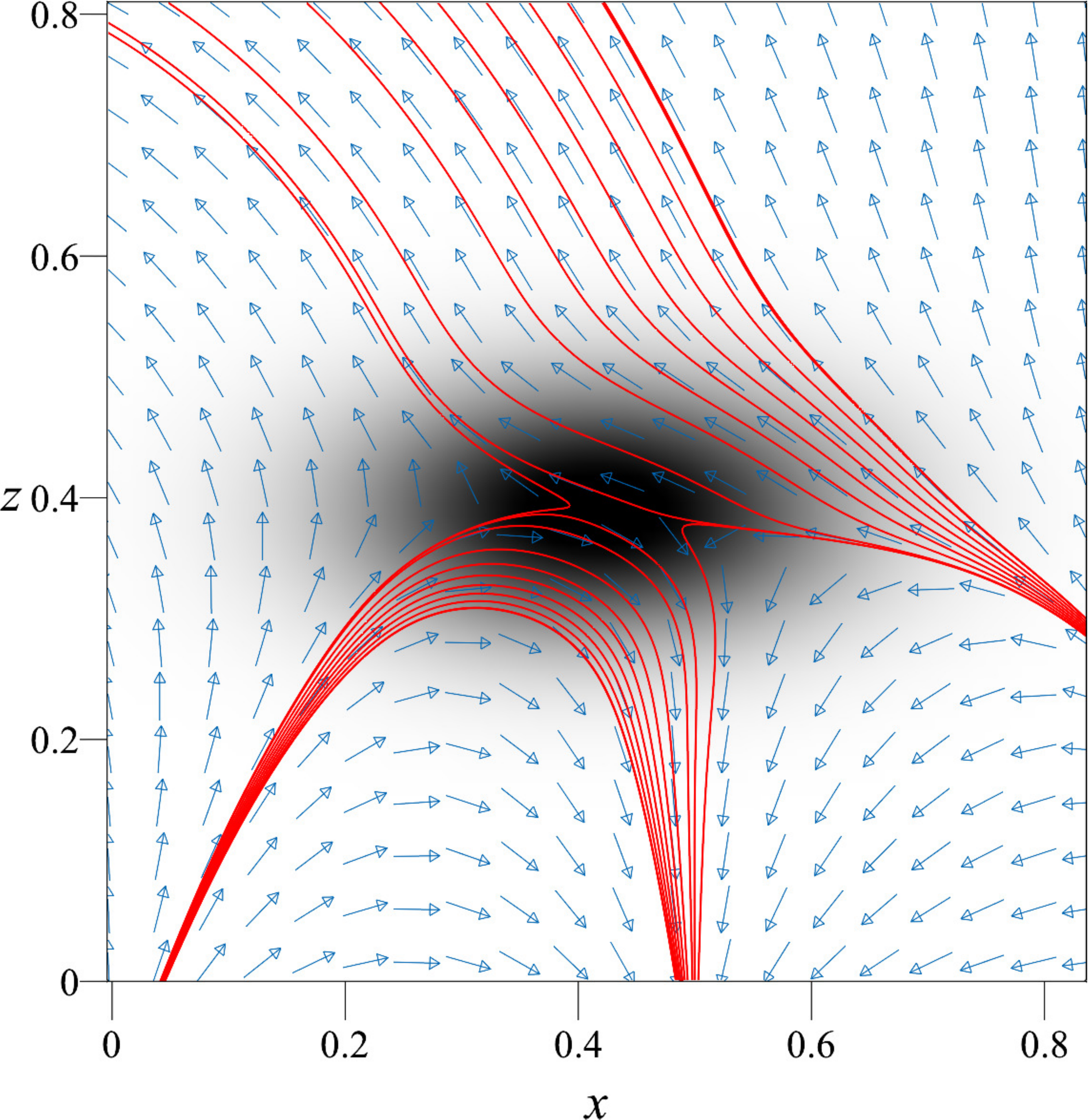}
\caption{Sample magnetic field lines in the $y=0$ plane showing the null point structure at (a) $t=0$ and (b) $t=0.8$. Vectors show magnetic field orientation and the background shading is proportional to $E_y$. The field lines in (a) and (b) are traced from the same fixed footpoint locations in the positive polarities near the fan.}
\label{toy_collapse}
\end{figure}
We proceed to simulate the effect of a spine-fan magnetic reconnection process at the coronal null point as follows. It is known from previous numerical simulations that such a reconnection event occurs when the spine and fan of the null collapse towards one another, and a localised region of parallel electric field forms around the null (associated with a current sheet there) parallel to the fan surface \citep{pontinbhat2007a}. We therefore introduce a perturbation to the magnetic field in the form of a ring of magnetic flux in the $xz$-plane, centred on the null, that leads to a collapsed spine-fan structure (see Figure \ref{toy_collapse}). This flux ring generates an electric current parallel to the fan surface, and, by choosing its modulus to grow linearly in time, we obtain a time-independent electric field, also parallel to the fan surface and localised around the null. Specifically, we take
\begin{equation}\label{b_ring}
\BB_j= b_0\, \frac{t}{\tau}\, \nabla\times\left(  \exp\left(-\frac{L}{4}(x-x_N)^2 - L(y-y_N)^2 - L(z-z_N)^2\right) {\bf e}_y\right),
\end{equation}
where $\xx_N=(x_N,y_N,z_N)$ is the location of the null, given above, and Faraday's law can be satisfied by taking
\begin{equation}
{\bf E}=\frac{b_0}{\tau} \exp\left(-\frac{L}{4}(x-x_N)^2 - L(y-y_N)^2 - L(z-z_N)^2\right) {\bf e}_y.
\end{equation}
This technique has also been used by \cite{wilmotsmith2011a}, who  investigated reconnection in the vicinity of a separator line. In what follows, we set $\tau=1$, $b_0=1.5\times 10^{-3}$ and $L=100$, and consider the effect of varying the time parameter $t$ between $t=0$ and $t=1$. The result of adding the flux ring of Equation (\ref{b_ring}) to the initial potential field (\ref{b_pot}) is to locally reduce the relative angle between the spine and the fan close to the null point -- see Figure \ref{toy_collapse}. This process of null collapse is a characteristic signature of this reconnection mode (see Section \ref{backsec}). If we were to continue to increase $t$ to values greater than 1 we would obtain a bifurcation of the coronal null point; by restricting to $t<1$ we ensure that at all times only a single null point is present, which remains at $\xx=\xx_N$ by symmetry.

\subsection{Flux transfer across the separatrix surface}
The perturbation magnetic field $\BB_j$ decays exponentially away from the null, and is effectively zero at the photosphere. This implies that $\BB$ at $z=0$ is independent of time. Crucially, however, the locations of the intersection of the spine and fan with the photosphere change in time, since the spine and fan field lines change their identities as a result of the local reconnection event at the null. The implication is that, since the separatrix surface moves in time, the quantities of flux connecting the different photospheric flux sources are also time-dependent. This is already clear from observing the field lines in the $y=0$ plane plotted in Figure \ref{toy_collapse}. These field lines are traced from fixed footpoints on the photosphere in the positive sources, i.e.~near the footprint of the fan dome. Locally closed field lines are seen to open by transferring across the separatrix surface, and vice-versa.

\begin{figure}[t]
\centering
(a)\includegraphics[height=4in]{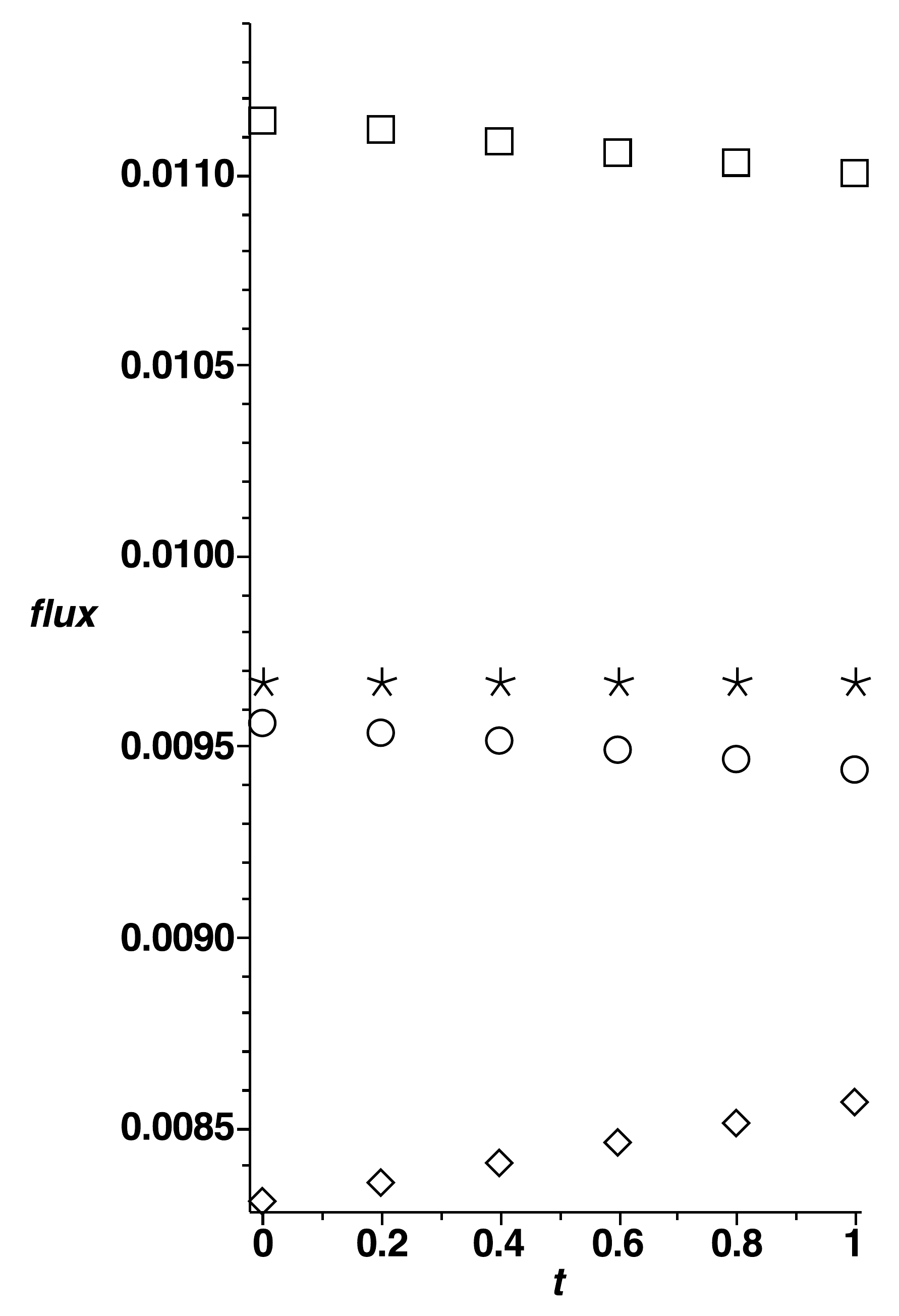}
(b)\includegraphics[height=3in]{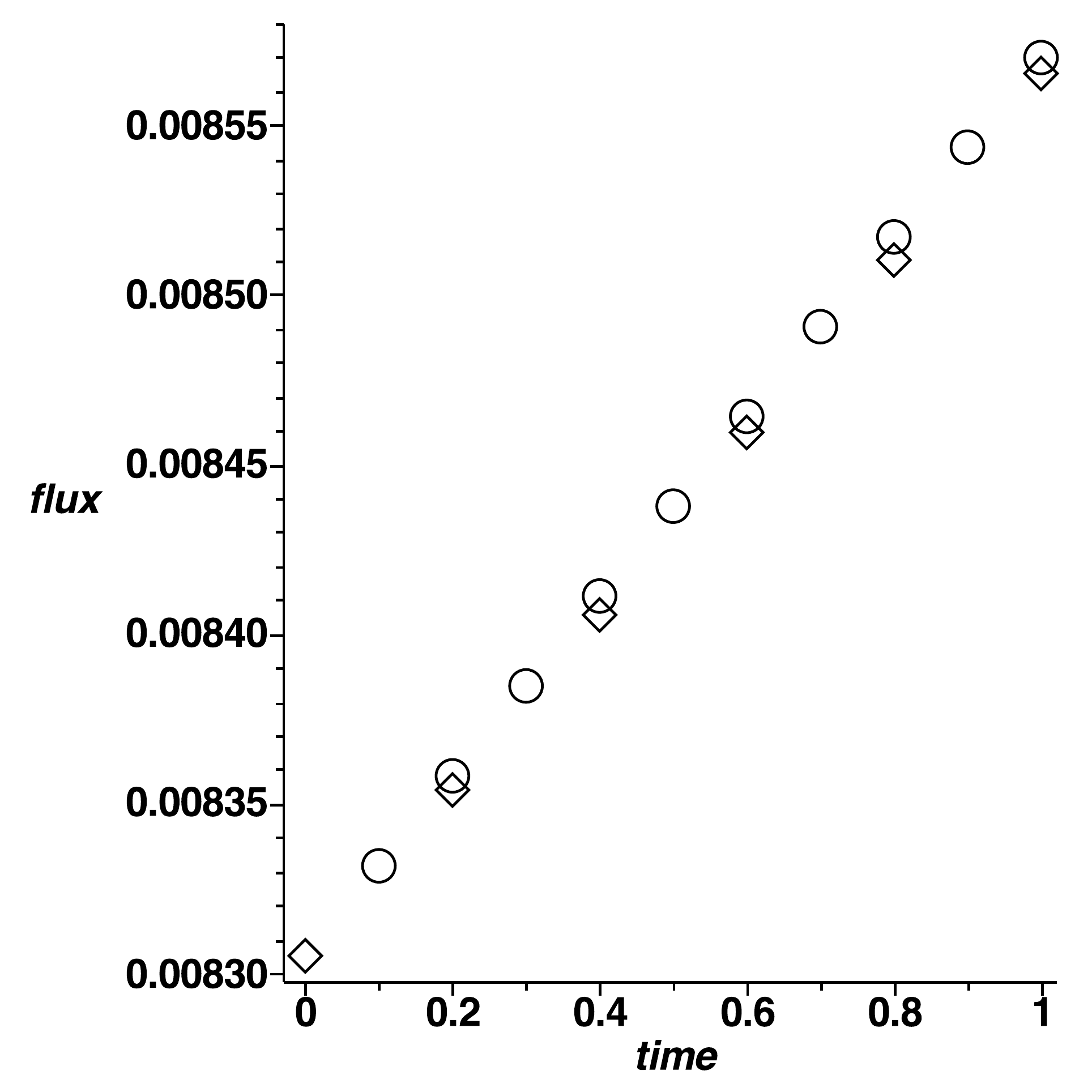}
\caption{(a) Magnetic flux as a function of time connecting the negative source 2 with positive sources 1 (diamonds), 3 (circles) and 4 (squares). Stars represent the total of these three values, which is constant. (b) Change of flux connecting source 1 to source 2, measured by integrating in time the reconnection rate given by Equation (\ref{recratedef}) (diamonds), and by performing a direct numerical integration of field lines and assessing their connectivity (circles).}
\label{toy_fluxes}
\end{figure}

We now calculate the change of magnetic connectivity induced by the simulated reconnection event, by calculating the total flux connecting the negative source 2 with each of the surrounding positive sources 1, 3 and 4. This is done by tracing field lines from 10000 starting points distributed across source 2 and assessing the locations at which they each return to the photosphere. The results are presented in Figure \ref{toy_fluxes}(a), from which it is clear that the flux connecting sources 1 and 2 increases in time while the flux connecting source 2 with sources 3 and 4 decreases with time. Clearly the net flux must remain the same -- being the total flux through the photosphere within source 2.

It was argued by \cite{pontinhornig2005} that for the spine-fan reconnection mode the rate of flux transport across the separatrix surface coincides with the reconnection rate defined within the framework of general magnetic reconnection, i.e. with the maximum value of integrated $E_\|$
along all field lines in the fan surface (see Section \ref{recsec}). This can be verified for our present configuration by seeking the maximum value of integrated $E_\|$, integrating this quantity in time, and then comparing with the numerically evaluated connectivities. In Figure \ref{toy_fluxes}(b) we see an excellent agreement between the time-integrated reconnection rate and the flux connecting sources 1 and 2 as a function of time. The small discrepancies result from {the numerical integration of field lines, the fact that we have a finite number of field lines (and therefore how the flux is associated with each of them),} and the numerical time integration of the cumulative reconnected flux from the reconnection rate.

\subsection{Field line motion: flipping}
\begin{figure}[t]
\centering
\includegraphics[width=2in]{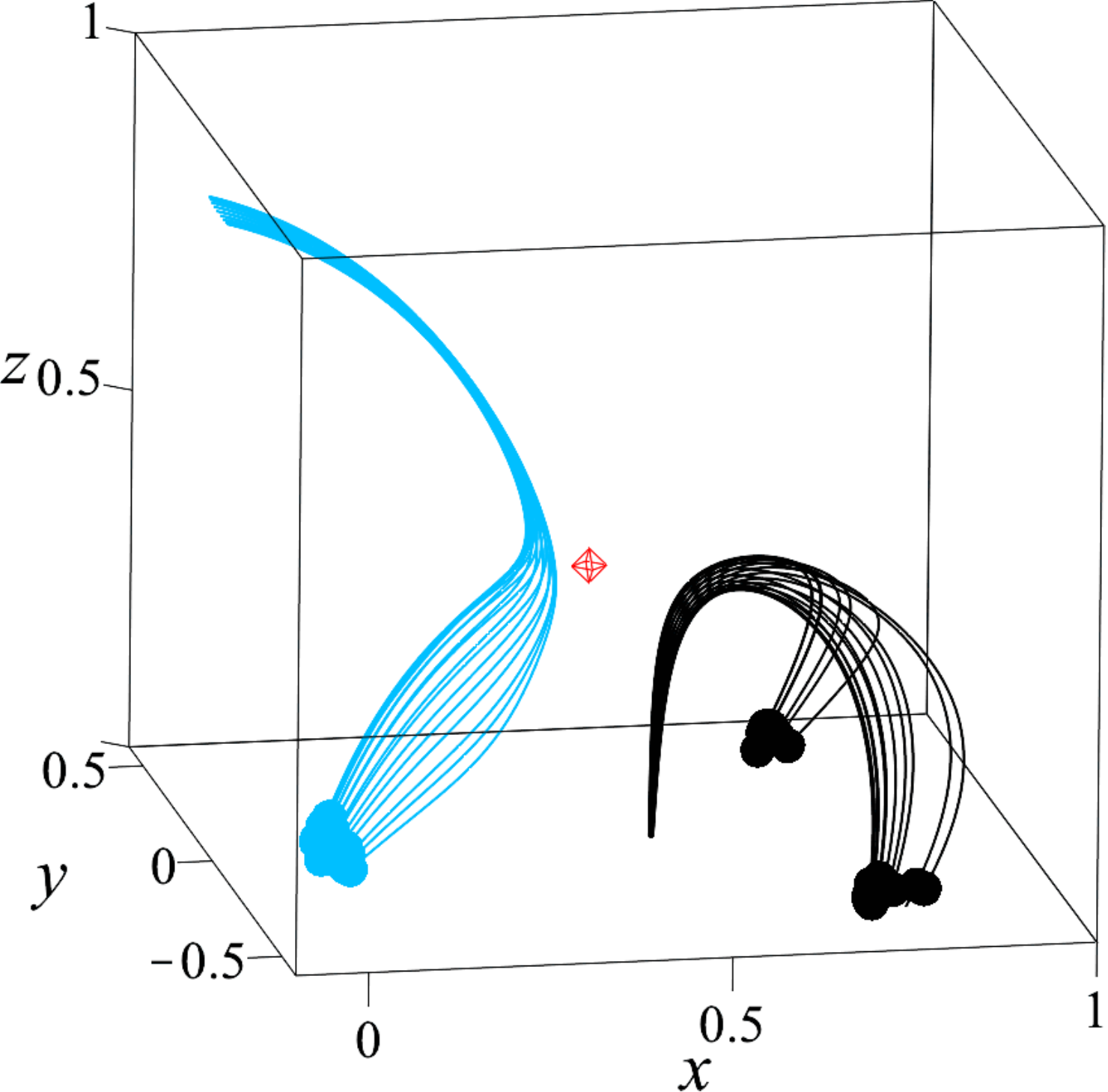}
~\includegraphics[width=2in]{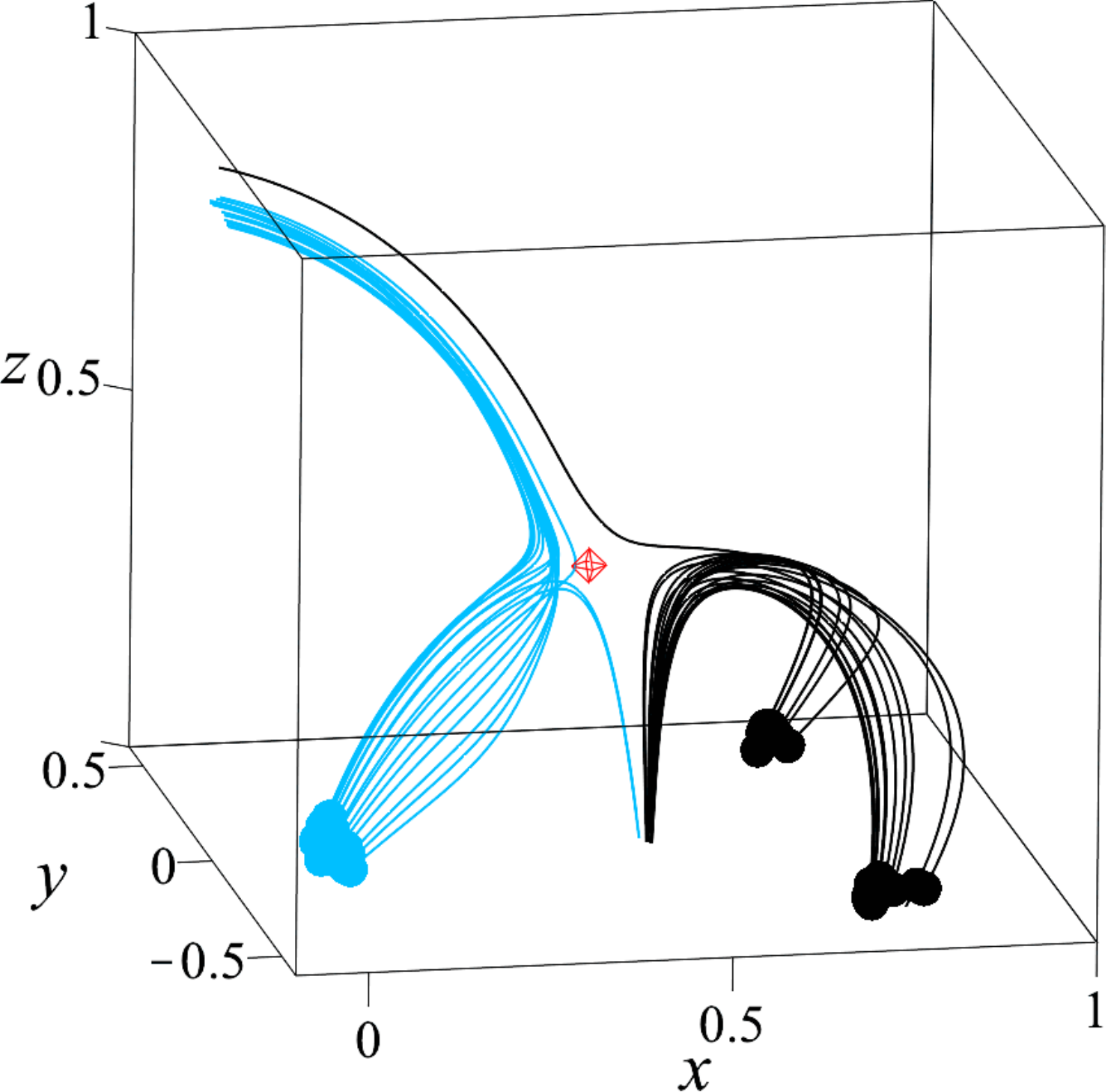}
~\includegraphics[width=2in]{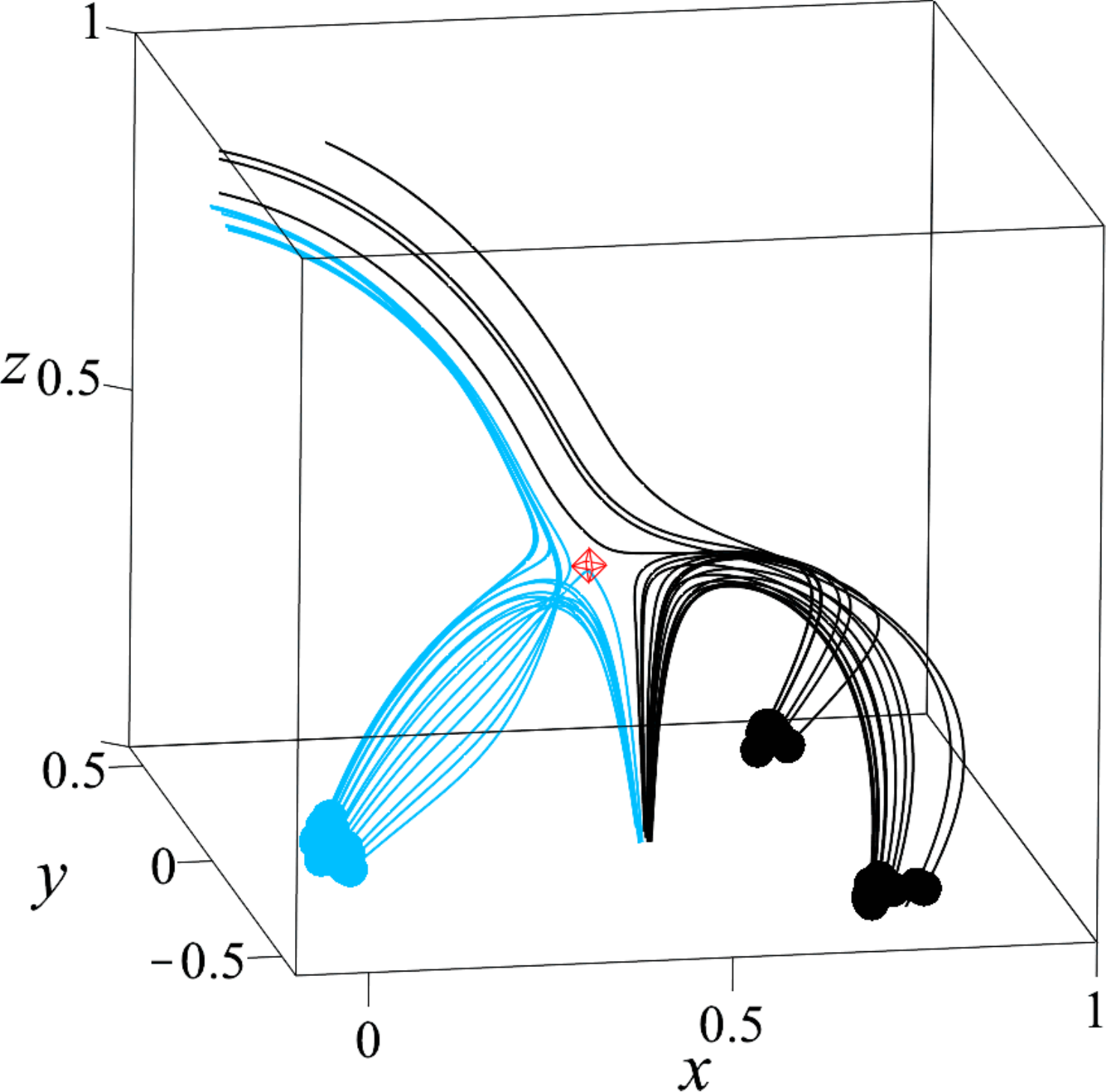}\\
~\includegraphics[width=2in]{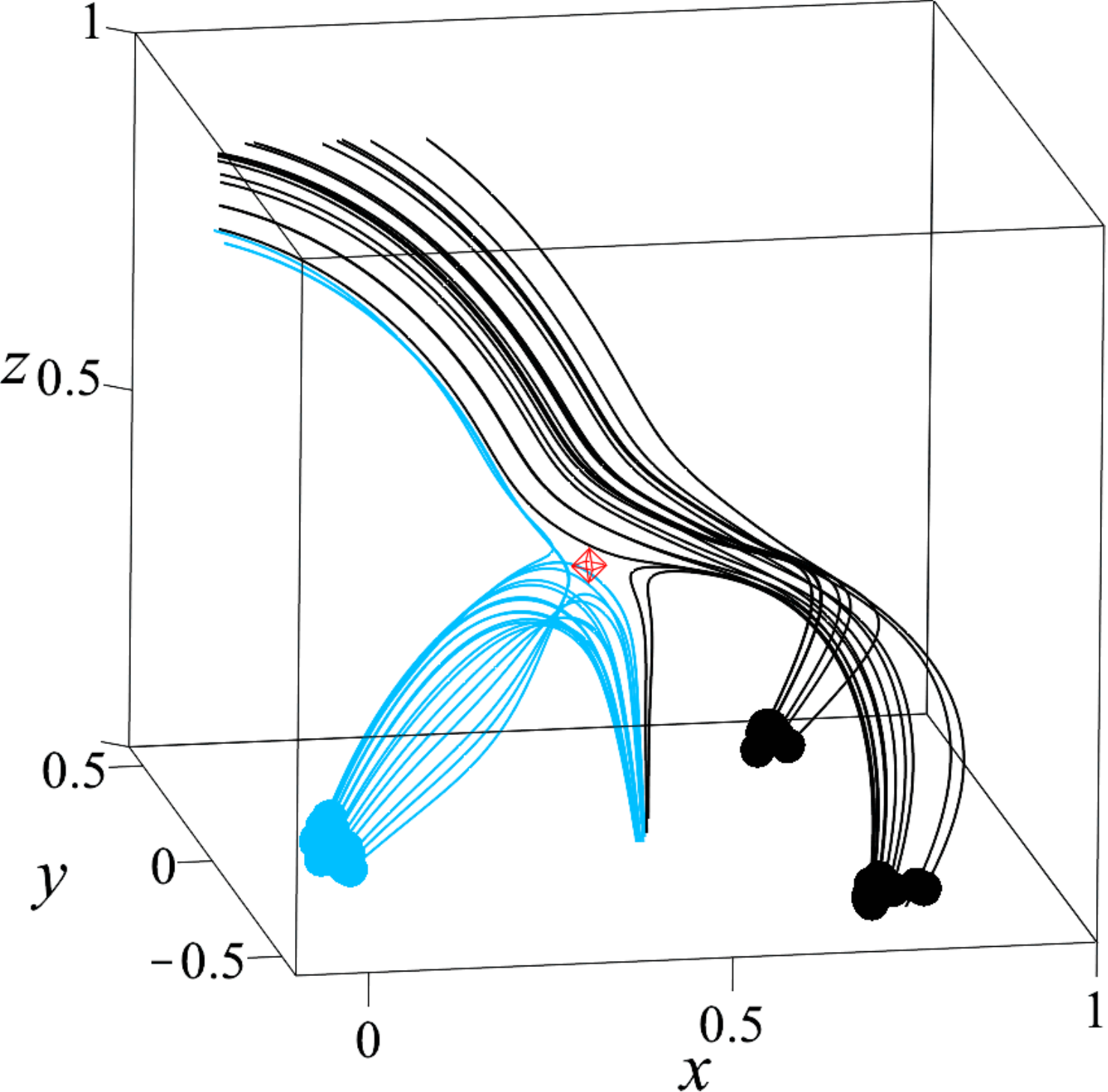}
~\includegraphics[width=2in]{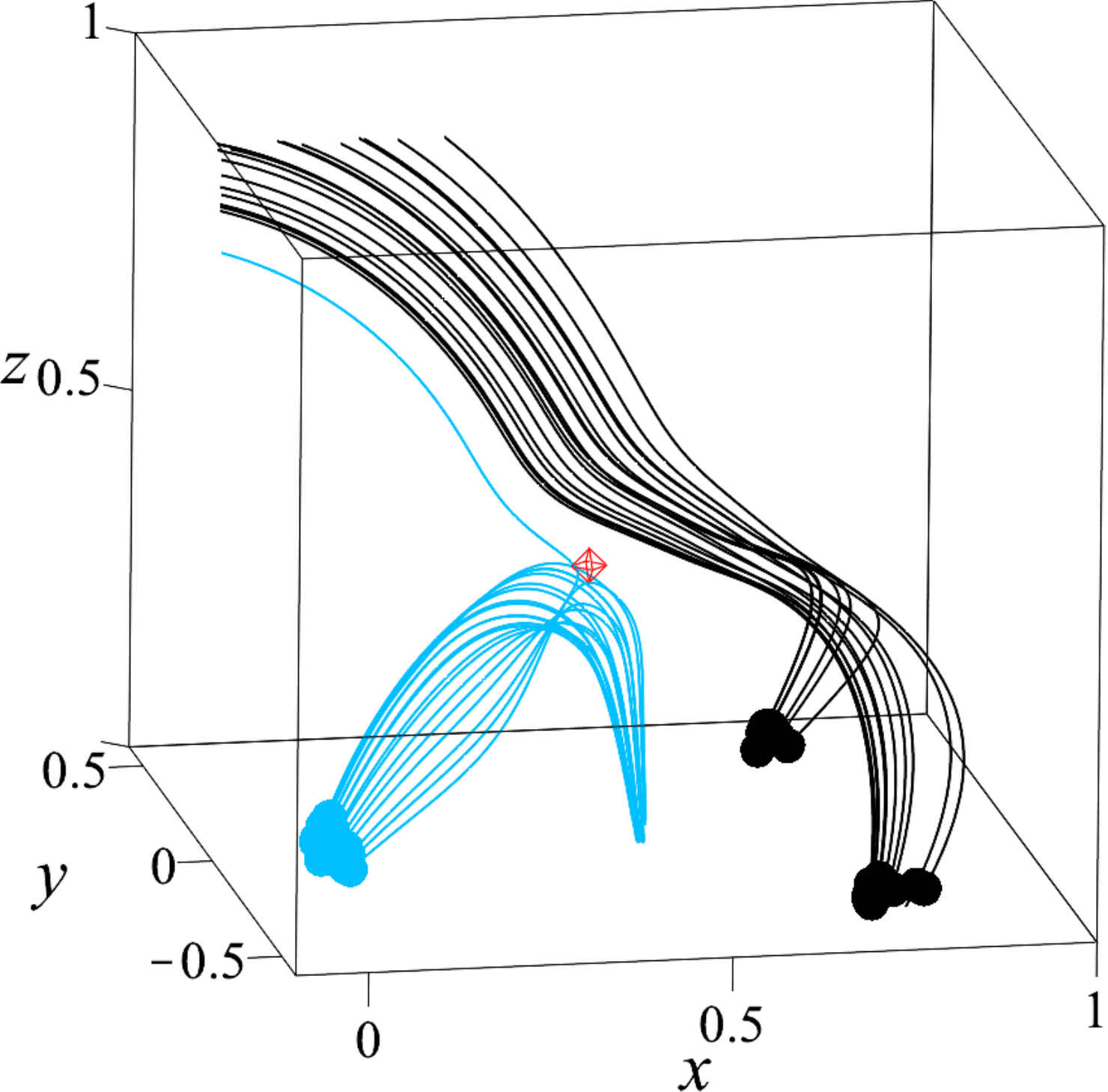}
~\includegraphics[width=2in]{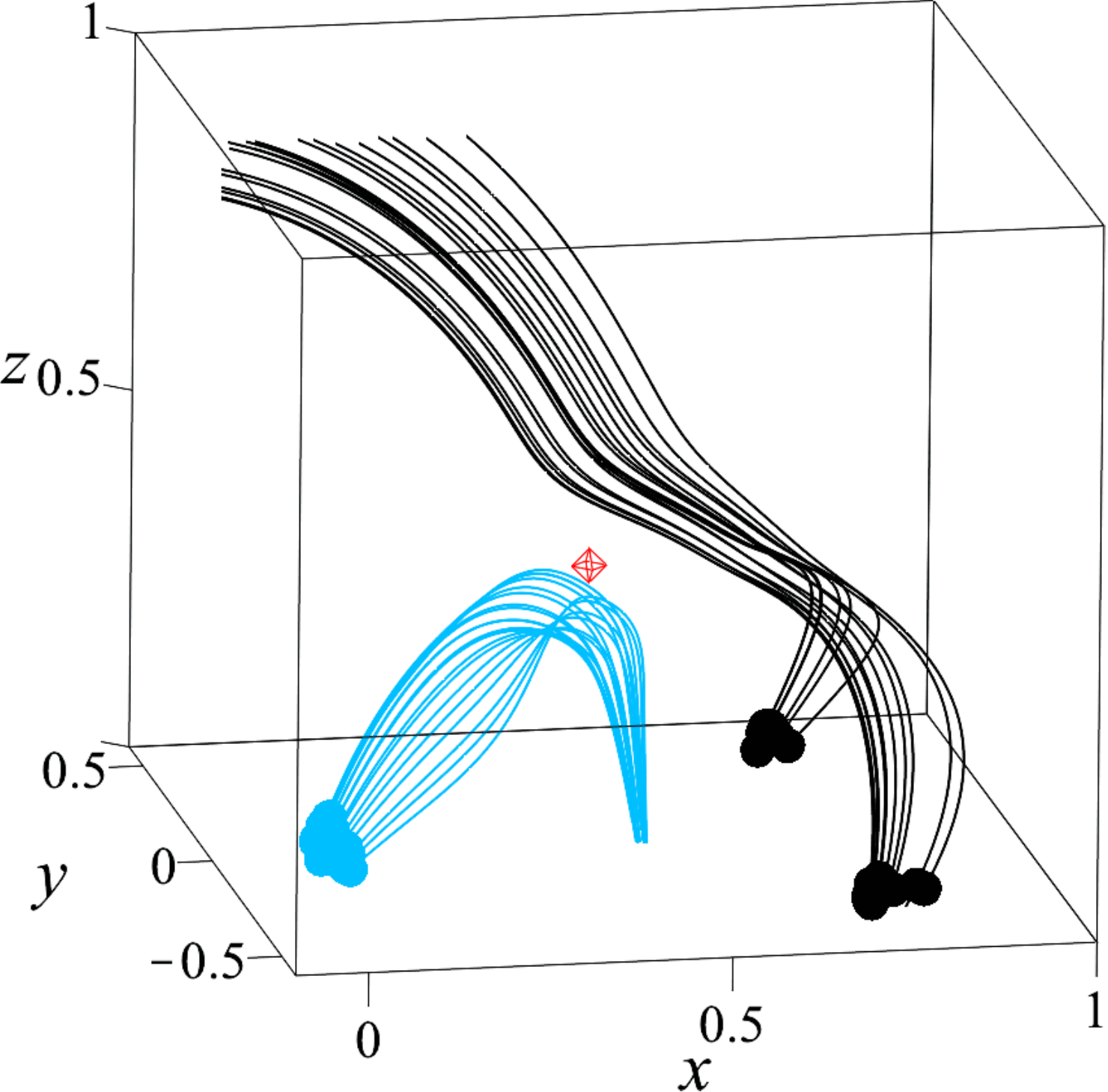}
\caption{Representative field lines traced from fixed footpoints (marked by spheres) located close to the footprint of the fan dome, at times $t=0,0.1,0.2,0.5,0.62,0.8$. The footpoints are located in the positive sources, so the motions of the field lines represent the flux velocity $\ww_{in}$. The location of the null is marked by the red diamond.}
\label{fliplines_fan}
\end{figure}
Three-dimensional reconnection involves a change of field line connectivity within a finite volume, very different from the two-dimensional case where field lines break only at the $X$-point (see Section \ref{recsec}). In 3D, in general, the field lines change connections continually and continuously as they traverse the non-ideal region (region within which $\EE\cdot\BB\neq 0$). That is, between any two adjacent times $t$ and $t+\Delta t$, {\it every} field line threading the non-ideal region changes its connectivity. This is a manifestation of the fact that no single flux transport velocity exists in the presence of a 3D-localised non-ideal region \citep[see][]{priest2003a}. In the case of spine-fan reconnection at a null point, this continual change of connections still occurs, with the field line mapping changing continuously for all field lines except those that pass instantaneously through the null itself, where the mapping is discontinuous.
\begin{figure}[t]
\centering
\includegraphics[width=2in]{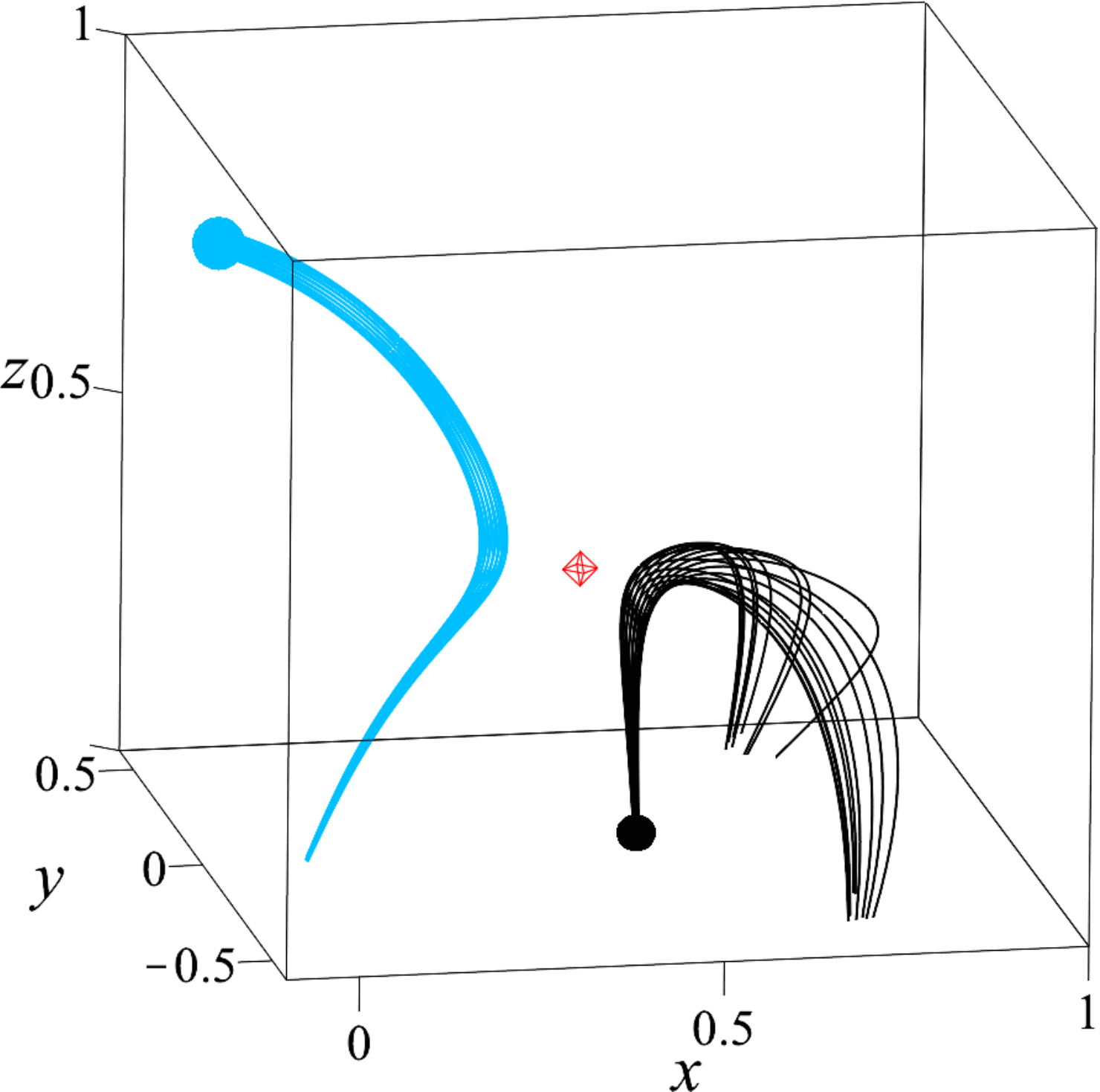}
~\includegraphics[width=2in]{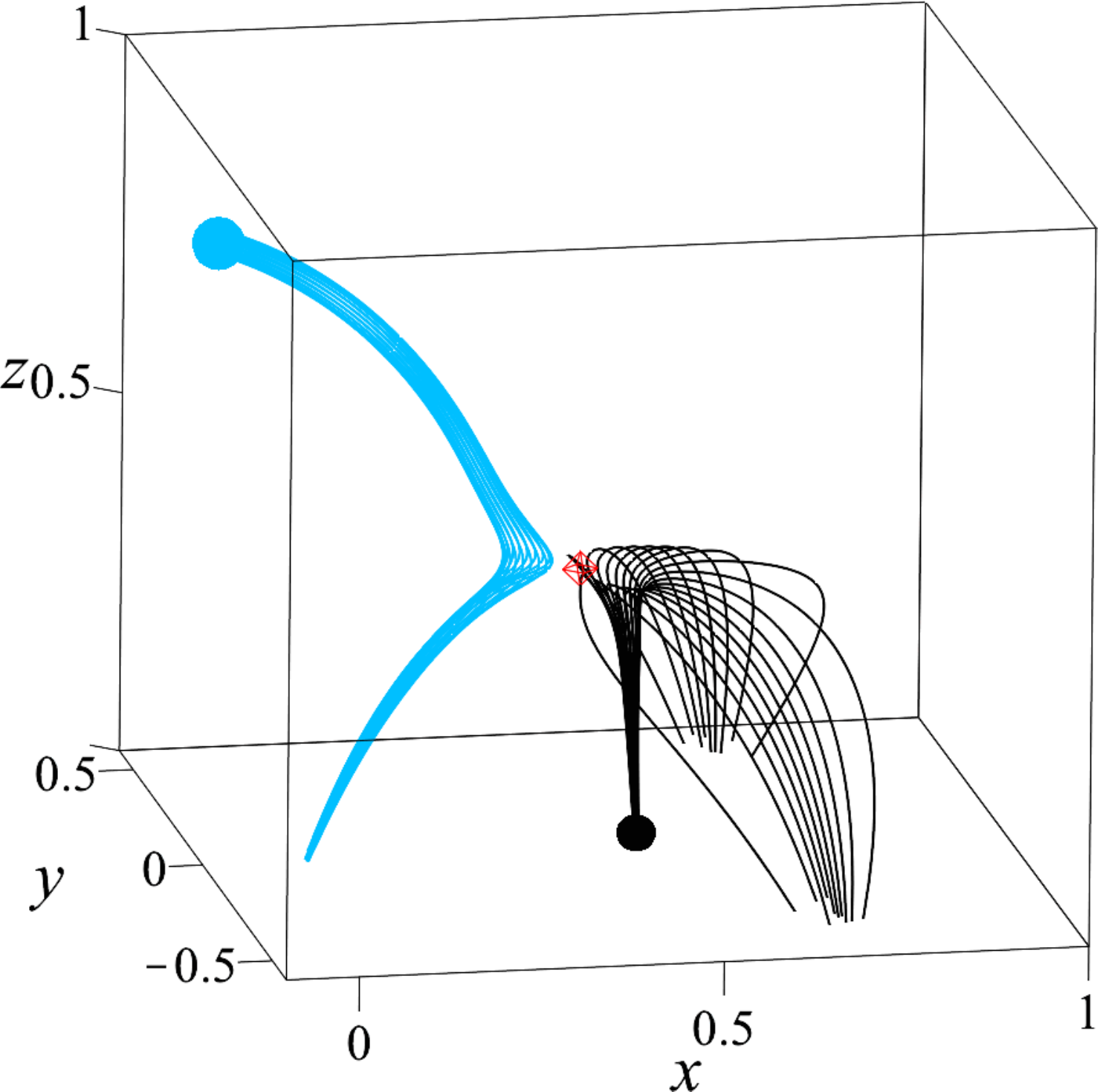}
~\includegraphics[width=2in]{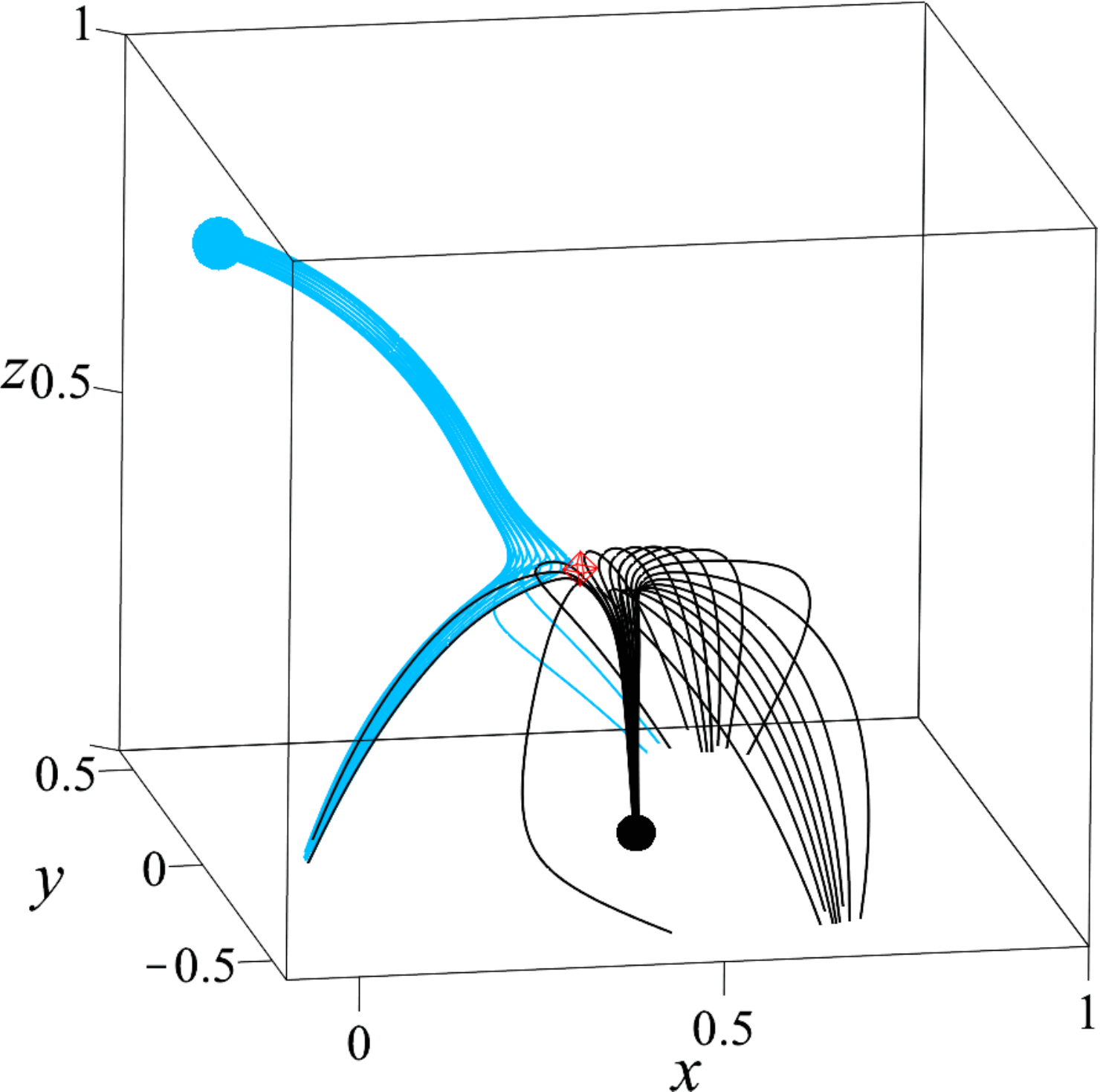}\\
~\includegraphics[width=2in]{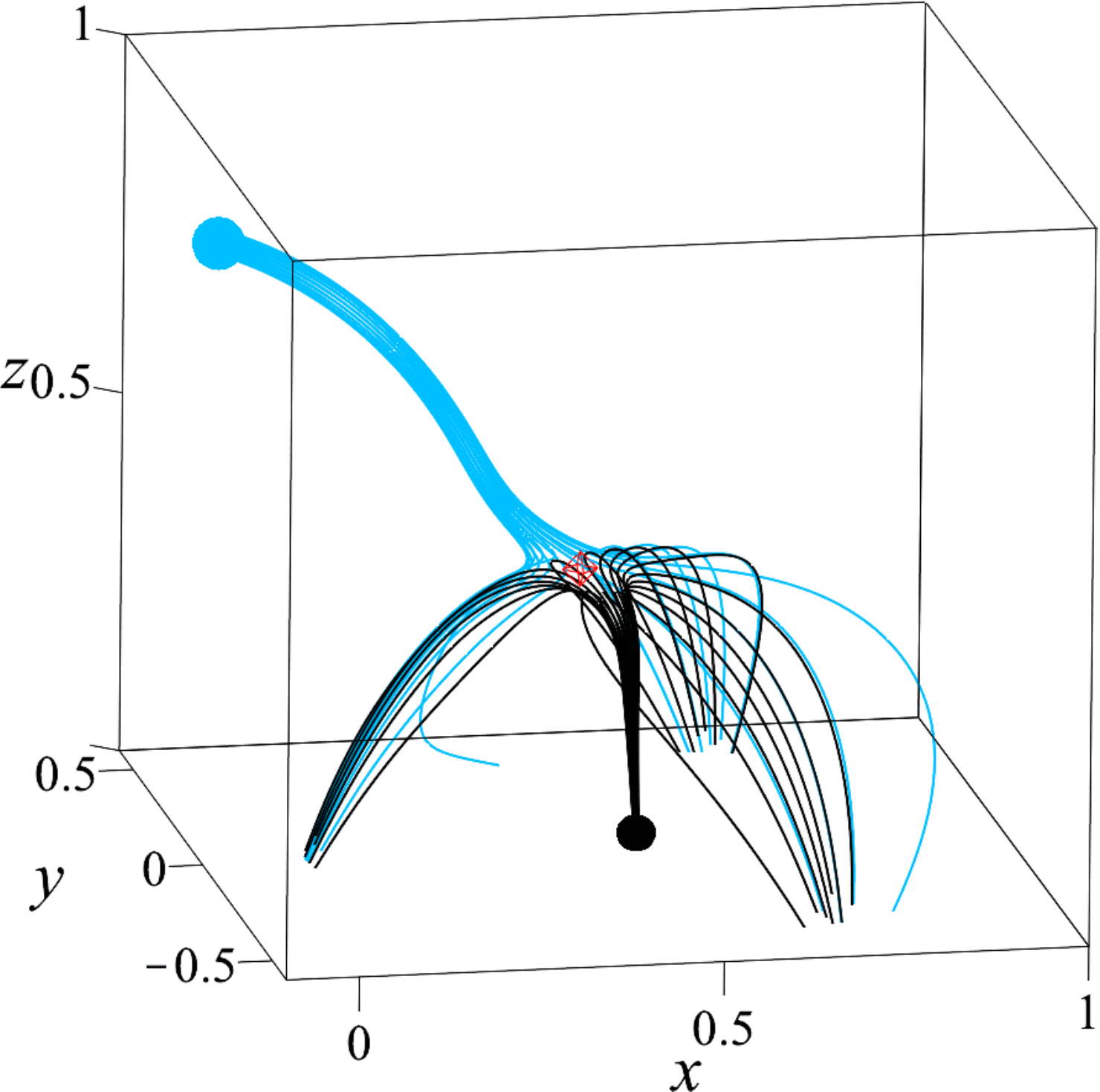}
~\includegraphics[width=2in]{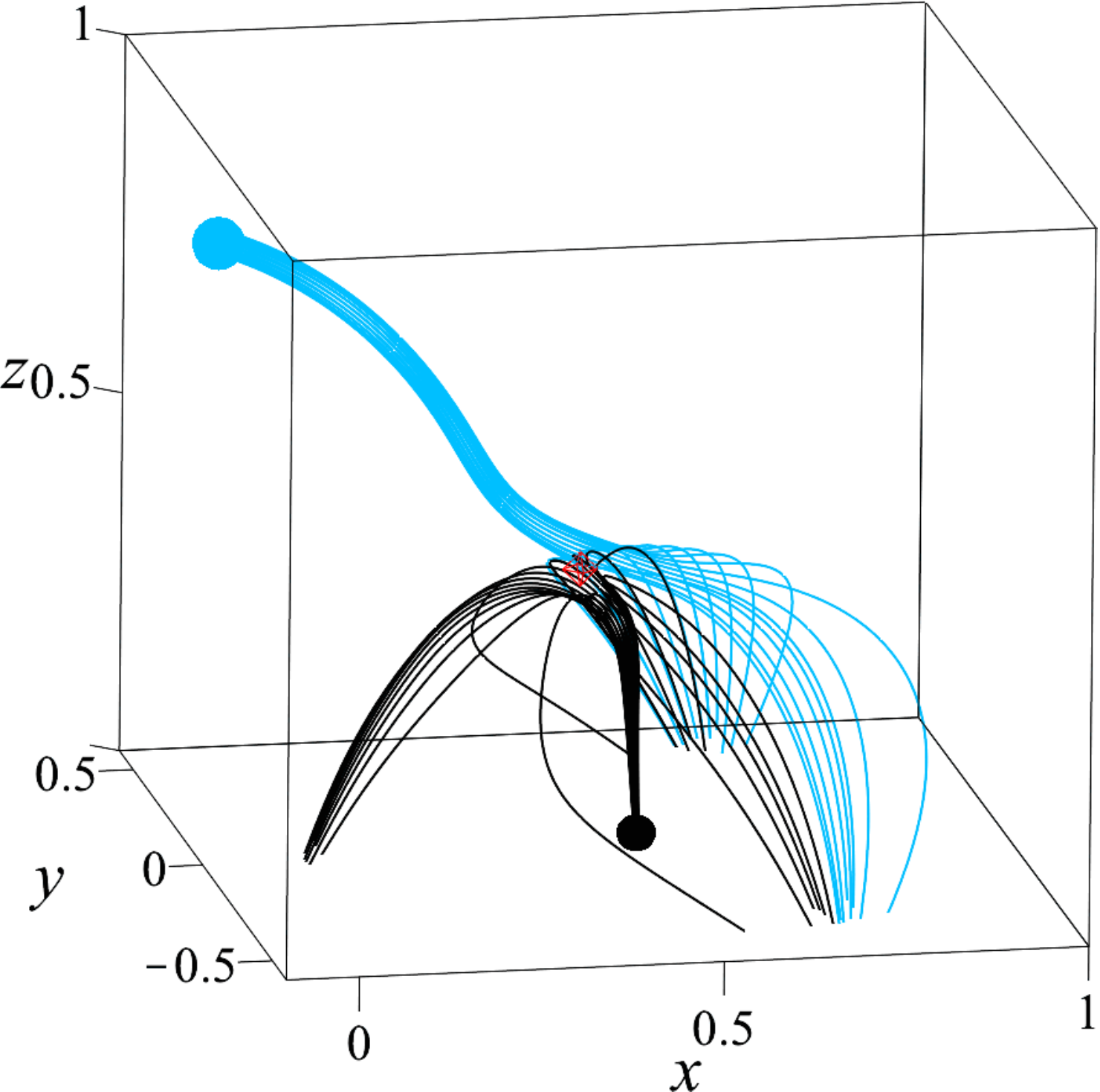}
~\includegraphics[width=2in]{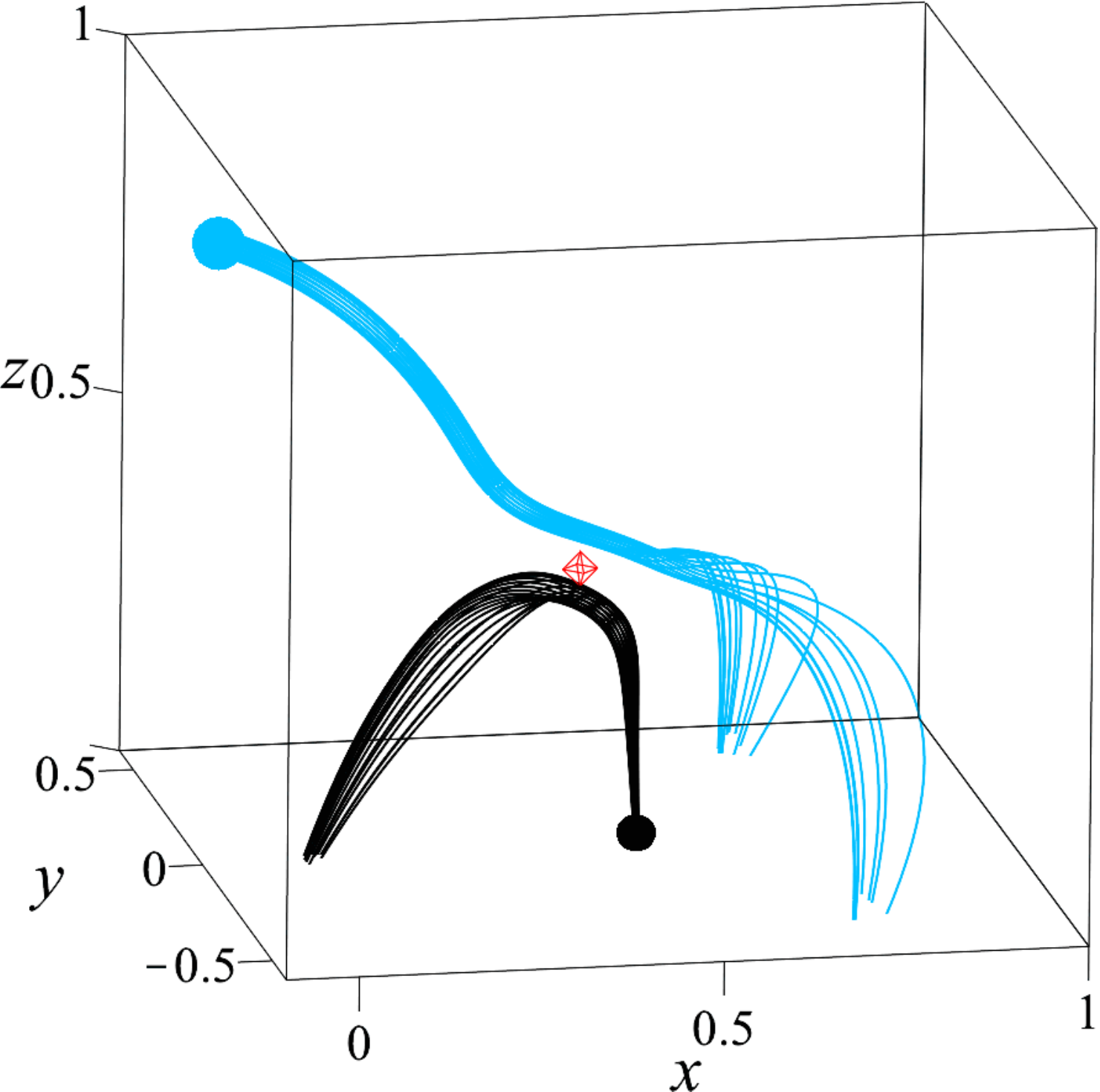}
\caption{Representative field lines traced from fixed footpoints (marked by spheres) located close to the spine axes, at times $t=0,0.33,0.38,0.5,0.62,0.8$. The footpoints are located in the negative sources, so the motions of the field lines exhibit the flux velocity $\ww_{out}$. The location of the null is marked by the red diamond.}
\label{fliplines_sp}
\end{figure}
\begin{figure}[t]
\centering
(a)\includegraphics[width=2.5in]{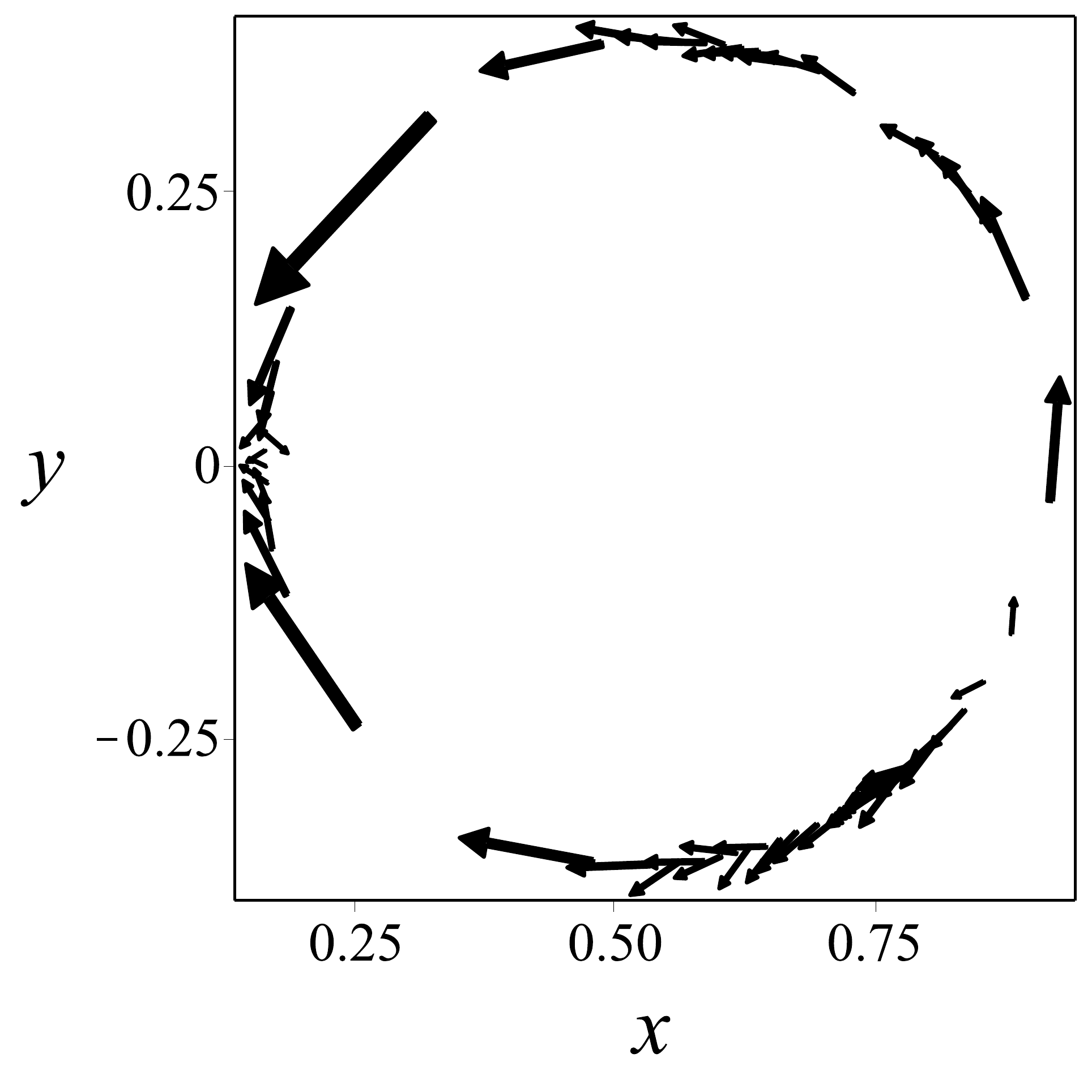}
(b)\includegraphics[width=2.5in]{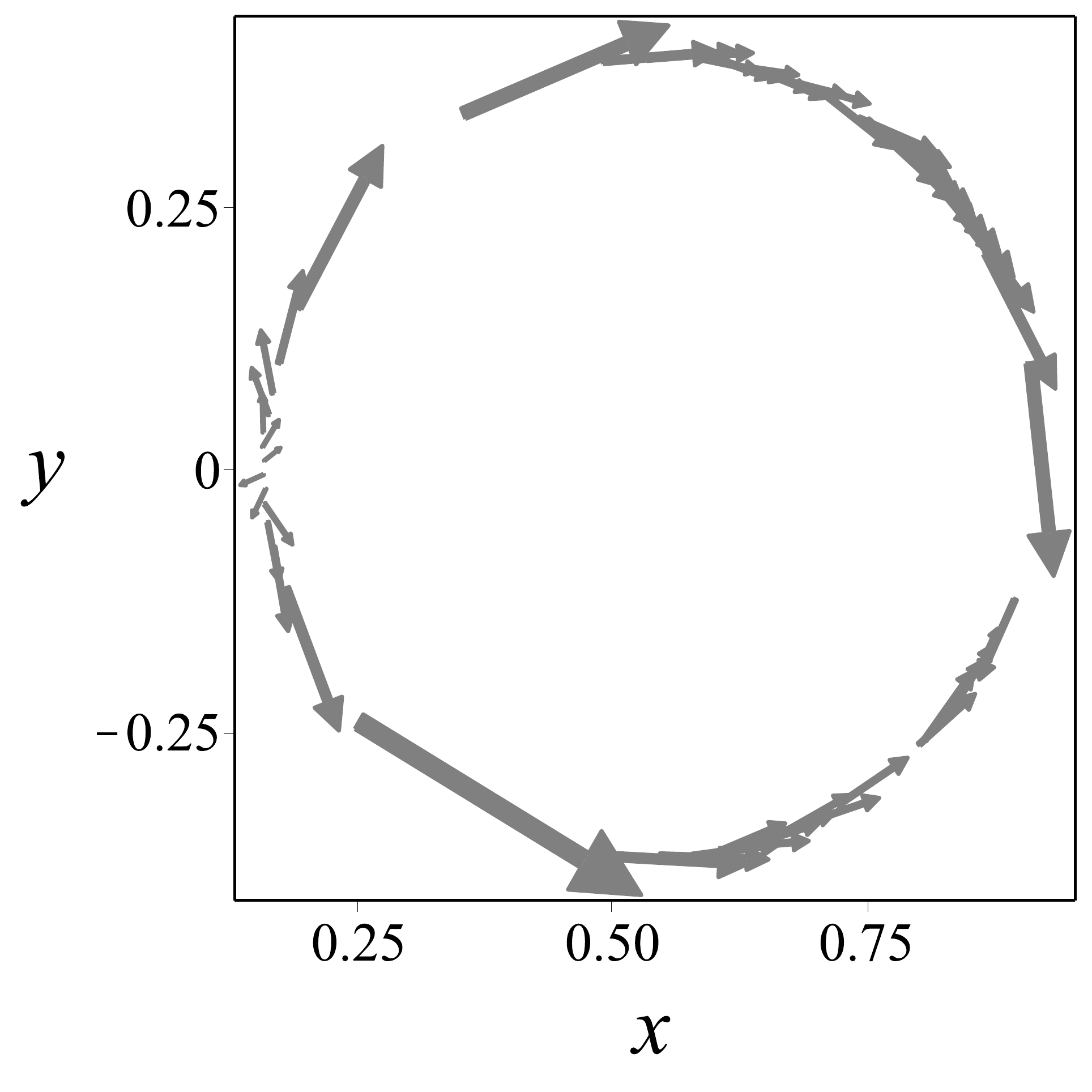}
(c)\includegraphics[width=2.9in]{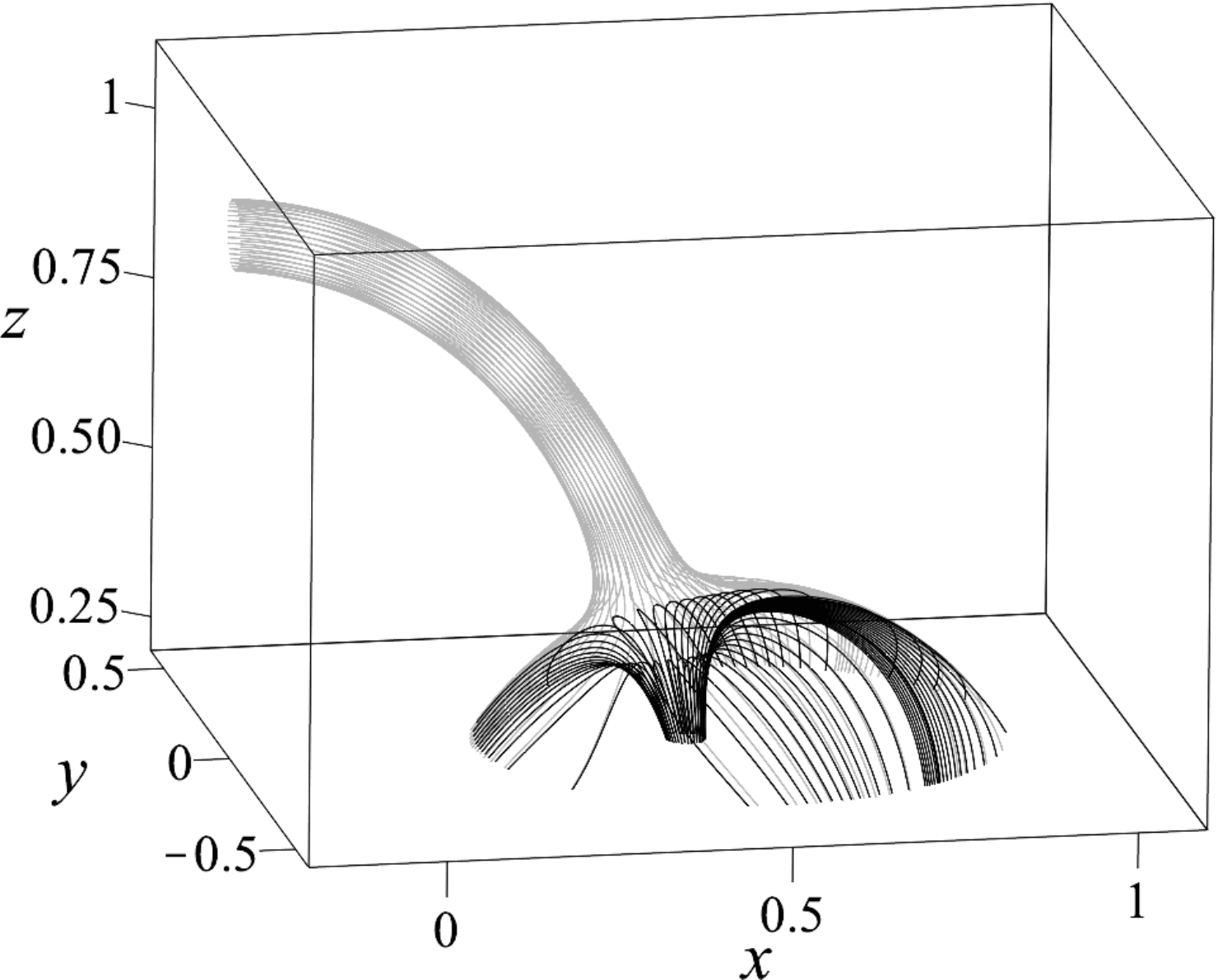}
(d)\includegraphics[width=2.1in]{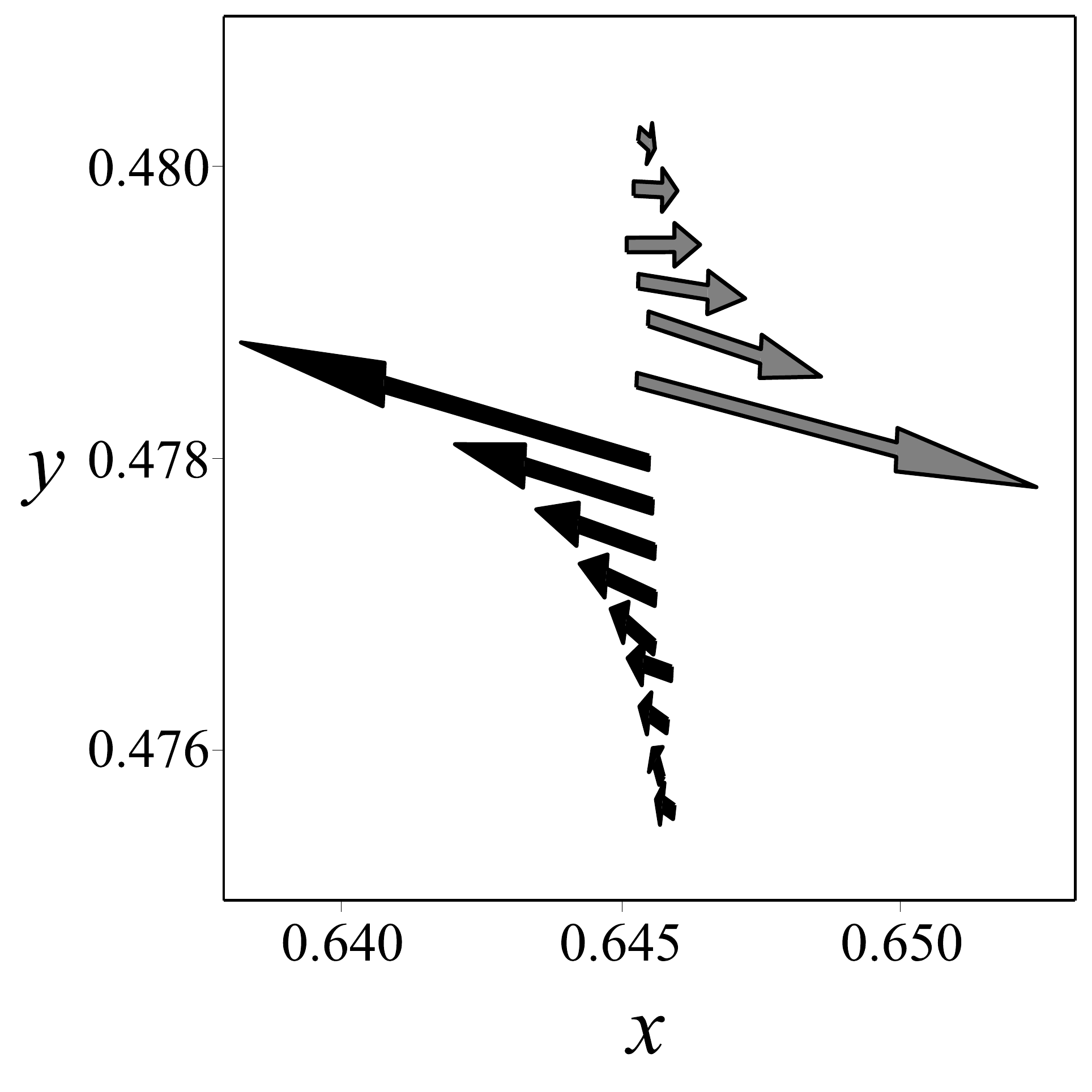}
\caption{(a, b) Maps of the flipping velocity ($\ww_{out}$) of field lines traced from points anchored in the negative flux patches close to the spine footpoints, at $t=0$, plotted on the plane $z=0.2$. The field lines are plotted in (c) -- black field lines lie inside the dome and have velocities shown in (a), whereas grey field lines lie outside and have velocities shown in (b). The arrows point in the direction of the flipping velocity, and their length is scaled with the square root of the velocity magnitude. (d) A close-up at a typical location around the separatrix surface footprint showing the velocity of field lines both inside and outside the separatrix dome, with the arrow length scaled to the velocity magnitude.}
\label{flipmap}
\end{figure}

The flux evolution for spine-fan reconnection about an (initially) linear null point has been described by \cite{pontinhornig2005} and \cite{pontinbhat2007a} in the kinematic and full MHD regimes, respectively. Here we describe how this translates to the separatrix dome geometry. Due to the non-existence of a single flux velocity, in order to visualise the field line evolution it is necessary to follow field lines anchored at either side of the non-ideal region independently. In 3D the field line velocity is in general not well defined, but we here follow previous work in making the physically motivated choice that in the ideal region field lines move at the {component of the plasma velocity perpendicular to ${\bf B}$} (which is in this case zero).

We first follow field lines from fixed footpoints anchored in the three positive sources close to the fan plane which move at velocity $\ww_{in}$ -- see Figure \ref{fliplines_fan}. We clearly observe the transfer of flux discussed above, with the blue field lines being transferred from outside to inside the separatrix surface, and black field lines being transferred from inside to outside. The field lines change connections continuously for most of their passage through the non-ideal region, both before and after the discontinuous jump in mapping at the separatrix surface.

Now consider field lines anchored in the two negative sources, close to the intersections of the spine with the photosphere. Some  representative field lines are plotted in Figure \ref{fliplines_sp}. As the field evolves the spine footpoints sweep across the sources, and field lines `flip' around from one side of the dome to the other, in opposite directions inside and outside the dome. We visualise the flipping velocity in Figure \ref{flipmap}. Here we trace a set of field lines lying close to the null (Figure \ref{flipmap}(c)) at $t=0$, anchored in the negative sources, and calculate where they move to some small time $\Delta t$ later. The associated $\Delta{\bf x}$, when divided by $\Delta t$, gives an approximate instantaneous velocity of field lines in the chosen plane. In order to improve the clarity of the figures we here represent the flipping velocity in the plane $z=0.2$ (note that the field perturbation is still negligible here). In Figures \ref{flipmap}(a) and \ref{flipmap}(b) the velocity is plotted for field lines lying inside and outside the separatrix dome, respectively. We clearly see the oppositely directed flipping from one side of the dome to the other. We also note that the flipping velocity is highest in regions where the field strength is lowest (required here for continuity). This flipping occurs in a thin envelope around the separatrix surface. In Figure \ref{flipmap}(d) we present a close-up at a typical location of the dome footprint. It is clear that the flipping velocity is fastest close to the separatrix surface, and falls away sharply as one moves away from the surface. Indeed, the flipping velocity must approach infinity as one approaches the separatrix surface, since there is one particular field line (at any given time) that by symmetry passes exactly through the spine and therefore has a discontinuous mapping, or infinite flipping velocity. Similarly, there are field lines that pass arbitrarily close to the spine (and therefore also the separatrix), and have arbitrarily large flipping velocities.

\section{Resistive MHD simulation}\label{numsec}
\begin{figure}[t]
\centering
(a)\includegraphics[height=3in]{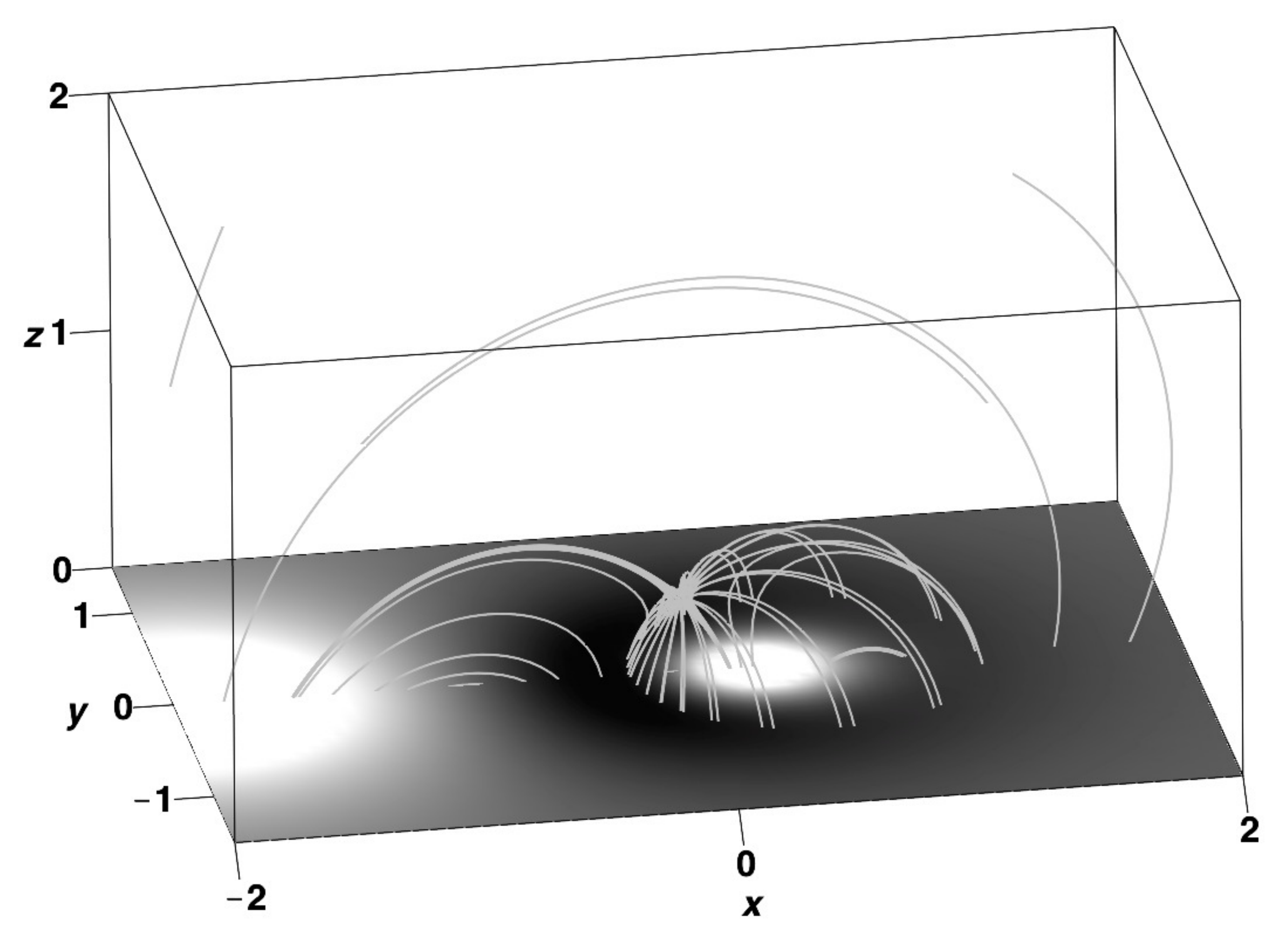}\\
(b)\includegraphics[width=1.8in]{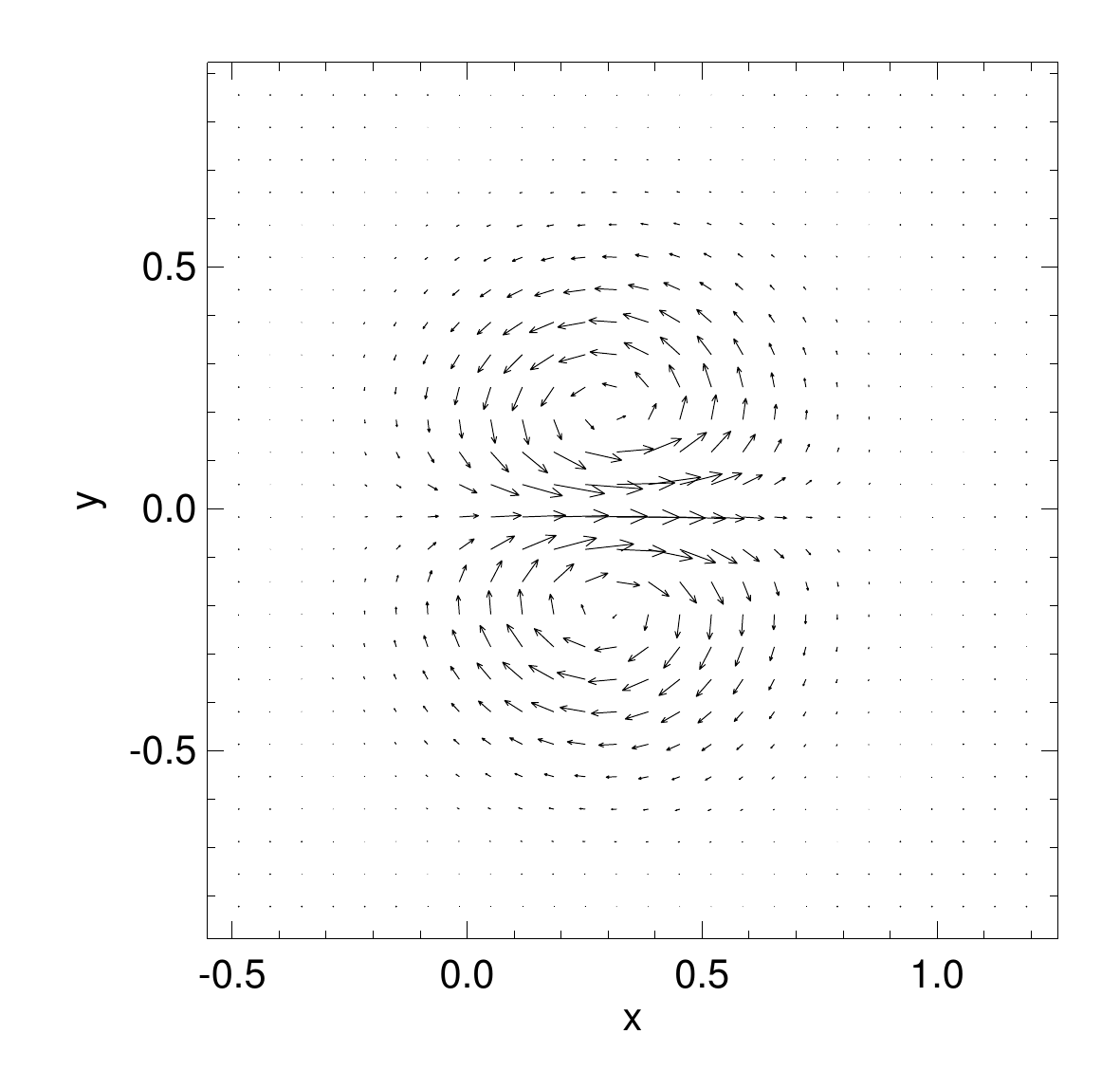}
(c)\includegraphics[width=1.8in]{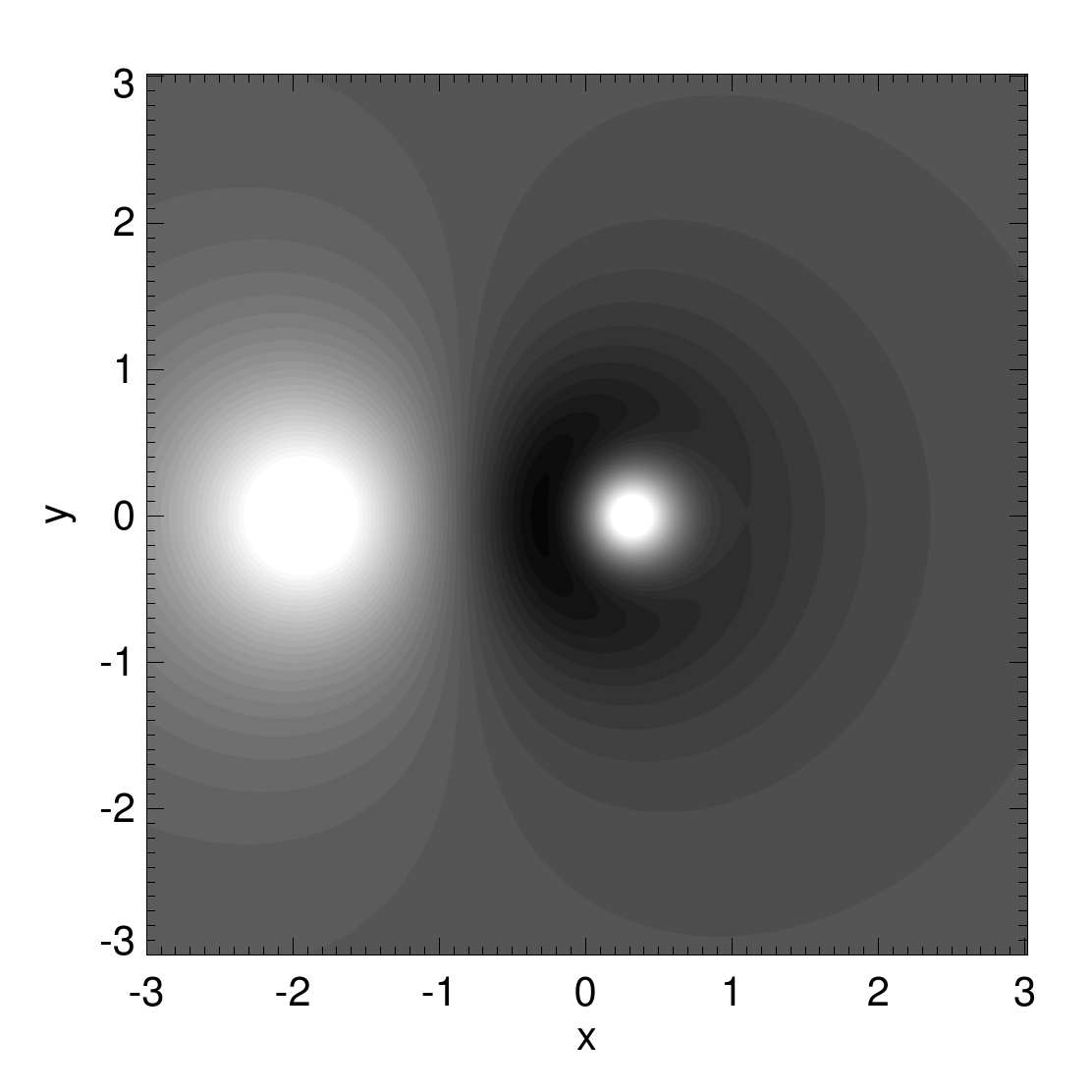}
(d)\includegraphics[width=1.8in]{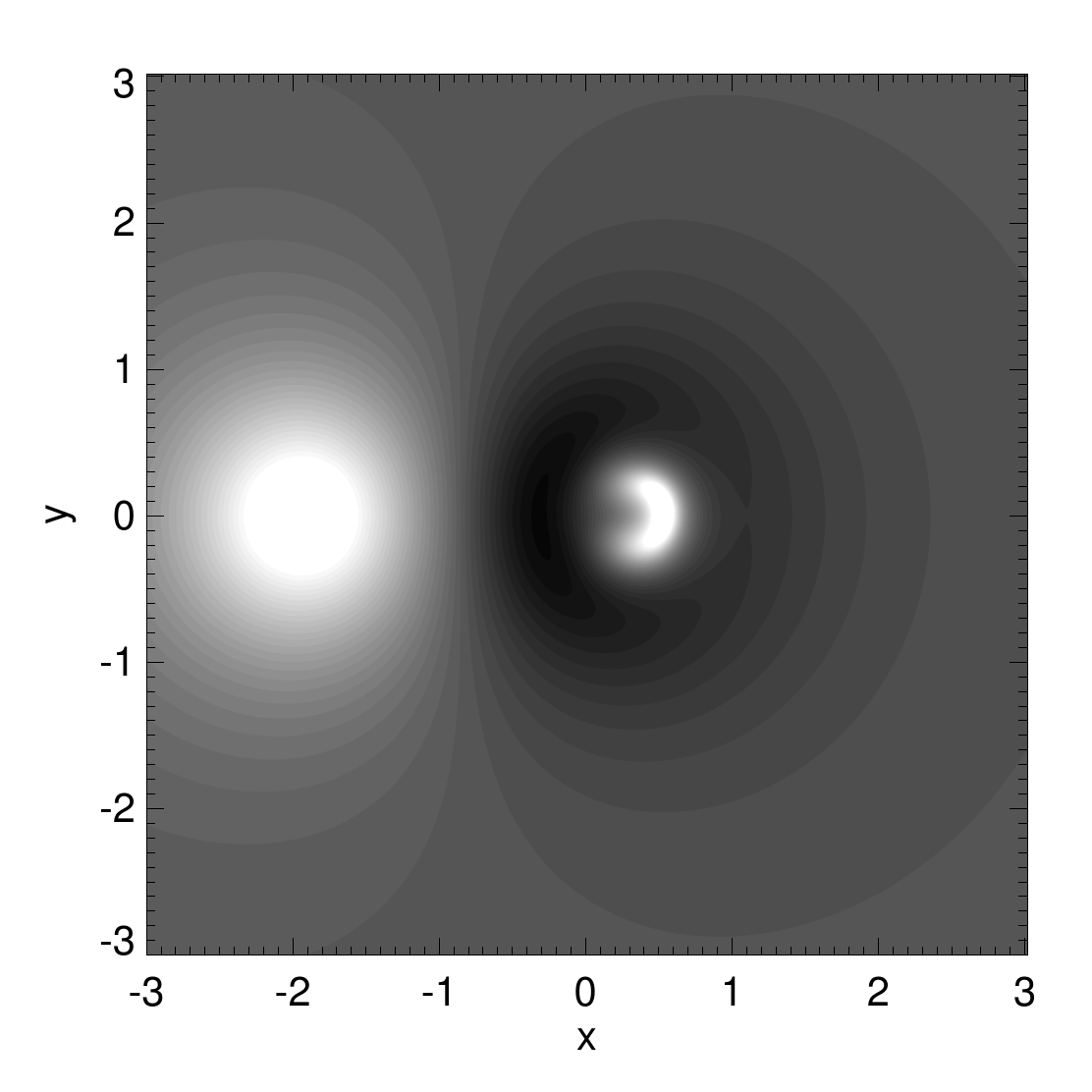}
\caption{(a) Representative magnetic field lines at $t=0$ in a subsection of the domain for the MHD simulation. The shading on the $z=0$ plane represents the vertical magnetic field strength on that plane. (b) Pattern of the driving flow close to the parasitic polarity. Also shown are the normal component ($B_z$) of the magnetic field at the photosphere, $z=0$, at (c) $t=0$ and (d) $t=3.0$.}
\label{setup_num}
\end{figure}

We now consider a full MHD evolution illustrating spine-fan reconnection at a coronal null point. We solve the resistive MHD equations numerically using the Copenhagen Stagger Code \citep{nordlund1997}. We use a grid of $360^3$ points distributed over $x\in[0,2],~ y,z\in[-3,3]$. All boundaries are line-tied, with ${\bf v}={\bf 0}$ everywhere except in a region of prescribed boundary driving at $z=0$ (see below). The density and pressure are initiated as spatially uniform ($\rho=1$, $p=0.05$), and a spatially uniform resistivity ($\eta=3\times 10^{-4}$) and viscosity ($\mu=3\times 10^{-3}$) are employed. The initial state is an equilibrium, with a potential magnetic field representing a bipole with an embedded parasitic polarity (at around $x=0.5$) above which is located a coronal null point, see Figure \ref{setup_num}.
The field has similar structure to the simple model described  in the previous section, but is easier to deal with numerically since the field strength around the separatrix surface is closer to being isotropic. It is 
generated in practice using  three point charges -- again outside the domain of interest. Specifically, ${\bf B}$ at $t=0$ is given by Equation (\ref{b_pot}) with $n=3$, $\{\epsilon_1,\epsilon_2,\epsilon_3\}=\{1.0,-1.0,0.3\}$, and $\xx_1=(-1.9,0,-1.0),~\xx_2=(0.1,0,-1.0),~\xx_3=(0.3,0,-0.4)$. This magnetic field has a structure similar to that associated with a C-class flare observed in active region AR10191, as studied by \cite{masson2009} and \cite{baumann2013}.

\begin{figure}[t]
\centering
\includegraphics[width=3in]{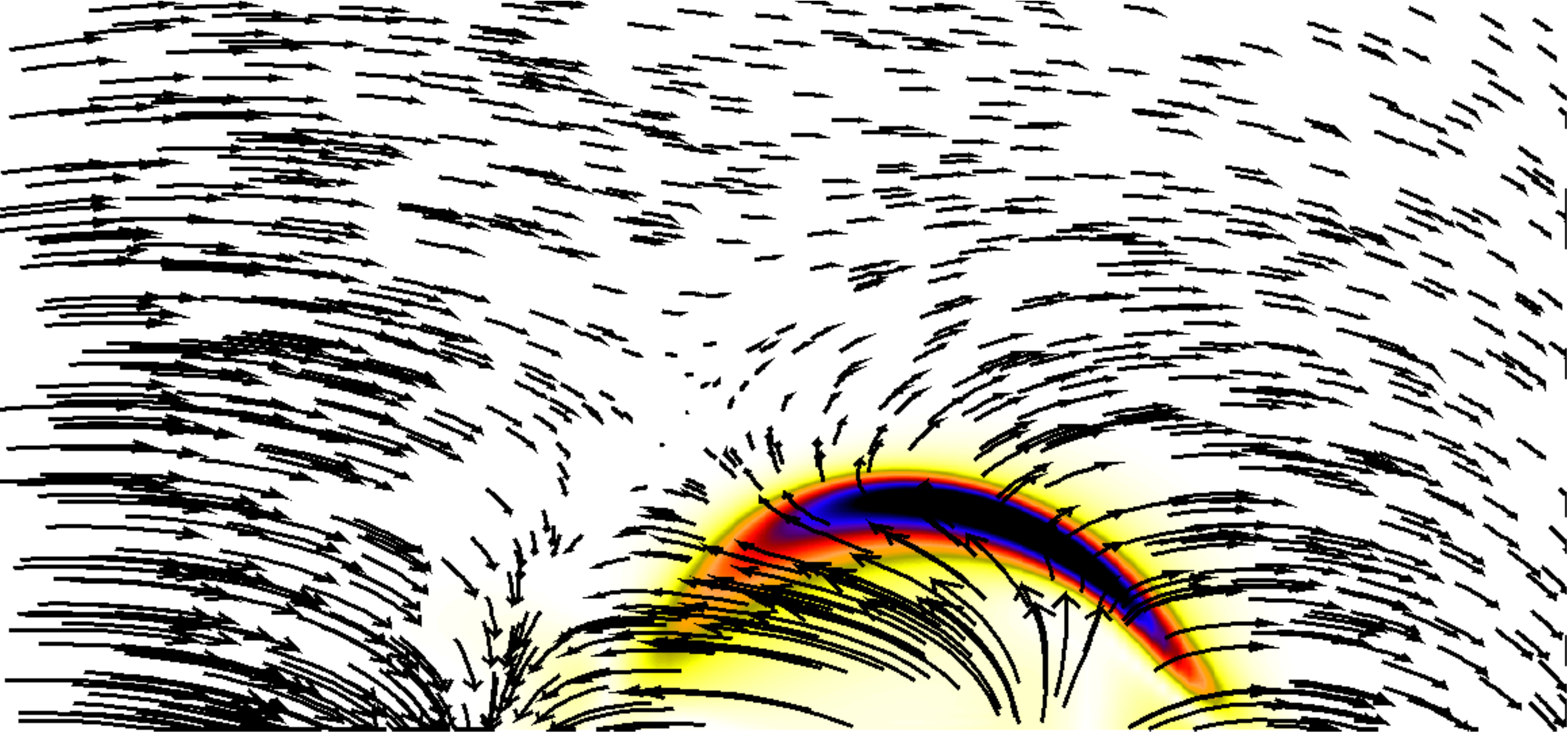}\\\vspace{0.3cm}
\includegraphics[width=3in]{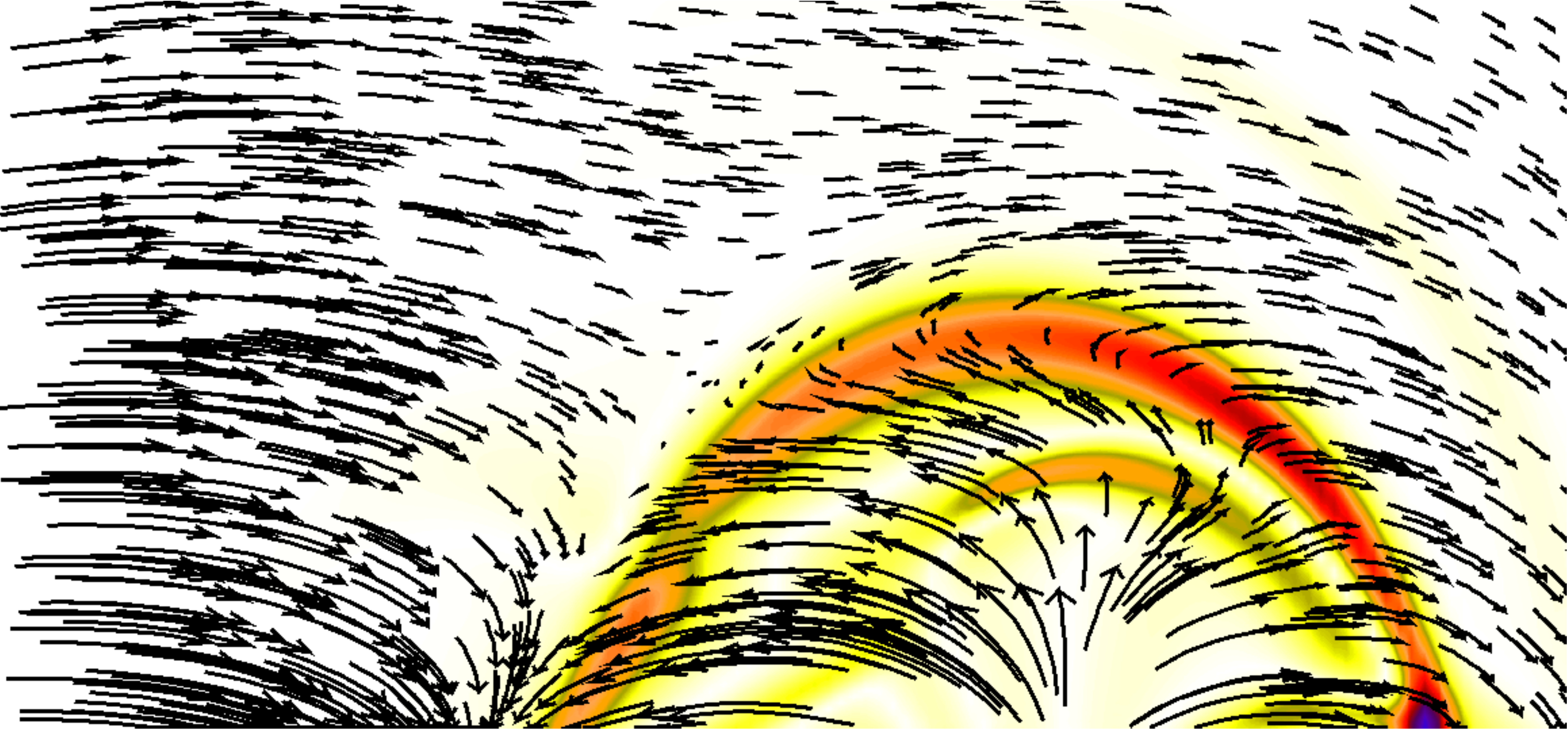}\\\vspace{0.3cm}
\includegraphics[width=3in]{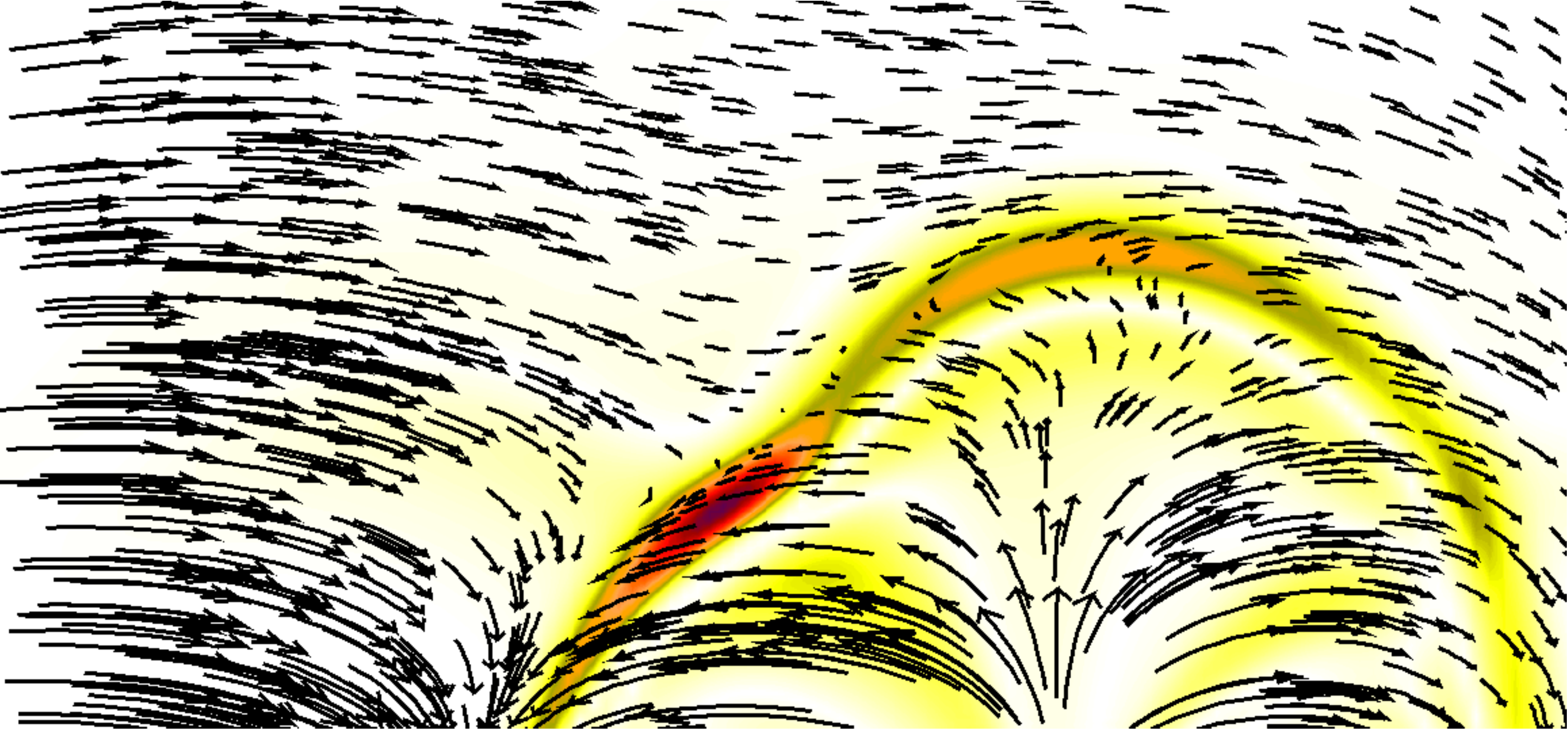}\\\vspace{0.3cm}
\includegraphics[width=3in]{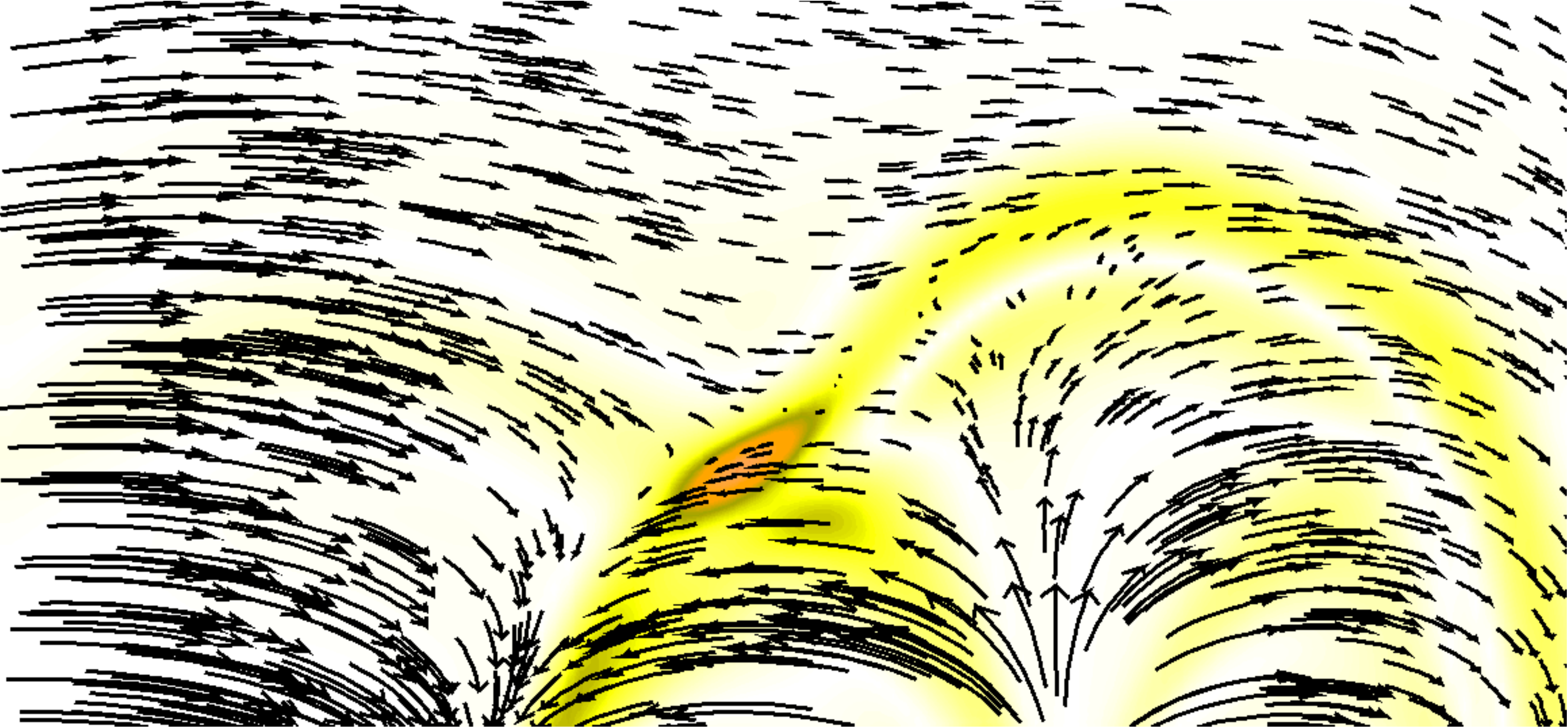}\\
\includegraphics[width=2in]{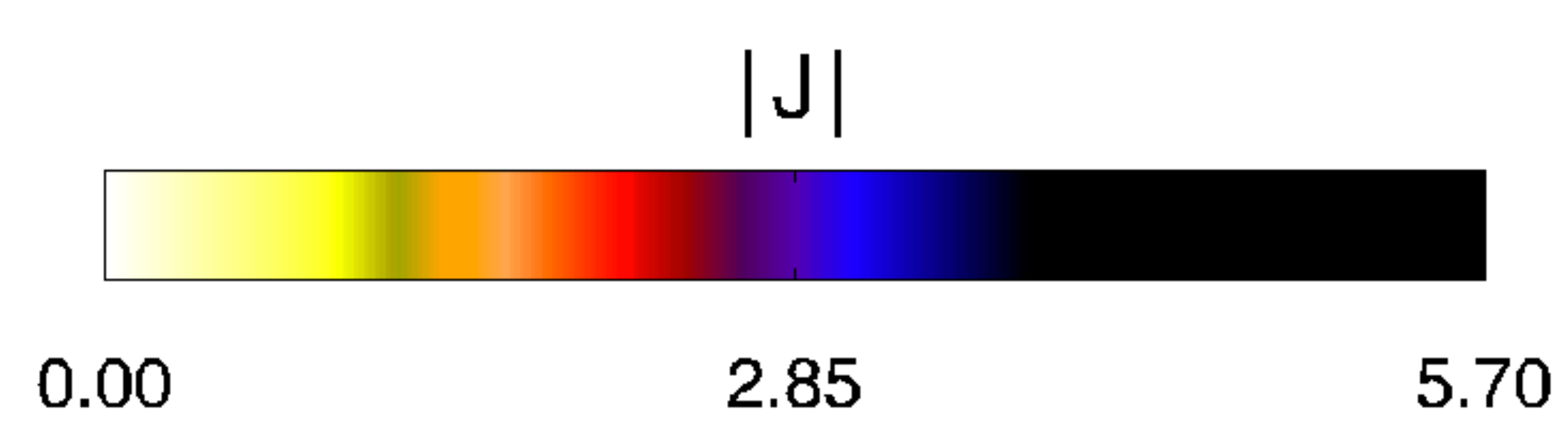}
\caption{Frames showing the magnetic field components (arrows) and current density (shading) in the $y=0$ plane over $x\in[-0.8,0.8]$, $z\in [0,0.8]$, for the MHD simulation. Top to bottom are at times $t=1.0, 2.4, 4.6, 6.0$.}
\label{jb_num}
\end{figure}

This equilibrium is disturbed by applying a boundary-driving velocity at $z=0$ that advects the parasitic polarity -- and associated spine footpoint -- in the positive $x$-direction.  In order to ensure that the flux through $z=0$ is preserved, this is done with an incompressible flow profile, so that the parasitic polarity becomes distorted (see Figure \ref{setup_num}). The driving velocity increases smoothly to a steady value, and then decreases smoothly to zero at $t=3$. As a result of this disturbance, a current front propagates upwards along the field lines beneath the separatrix and focuses eventually on the separatrix surface, in the vicinity of the null point. After the driving ceases, this current slowly dissipates (see Figure \ref{jb_num}).

\begin{figure}[t]
\centering
(a)\includegraphics[height=3in]{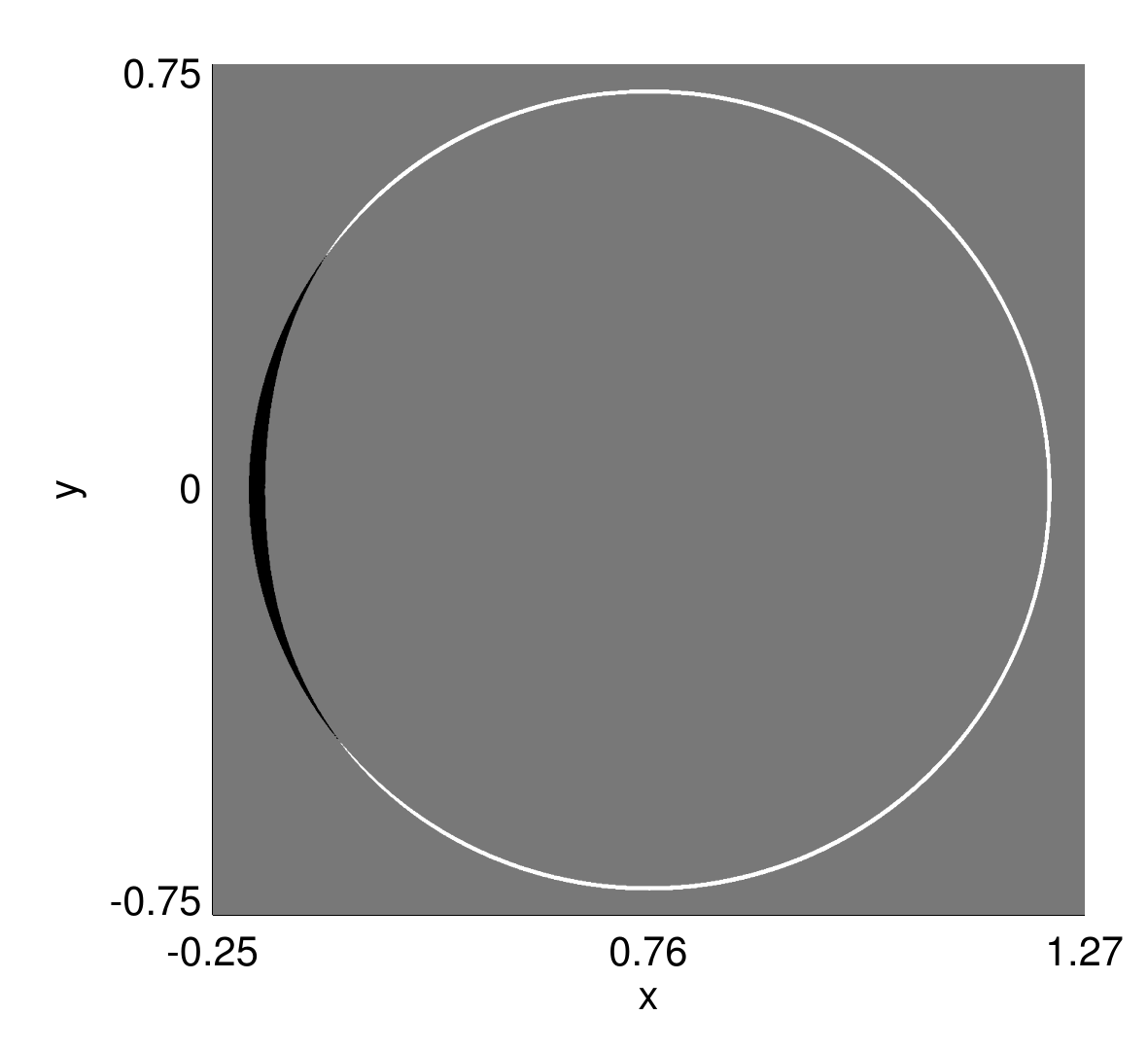}
(b)\includegraphics[height=3in]{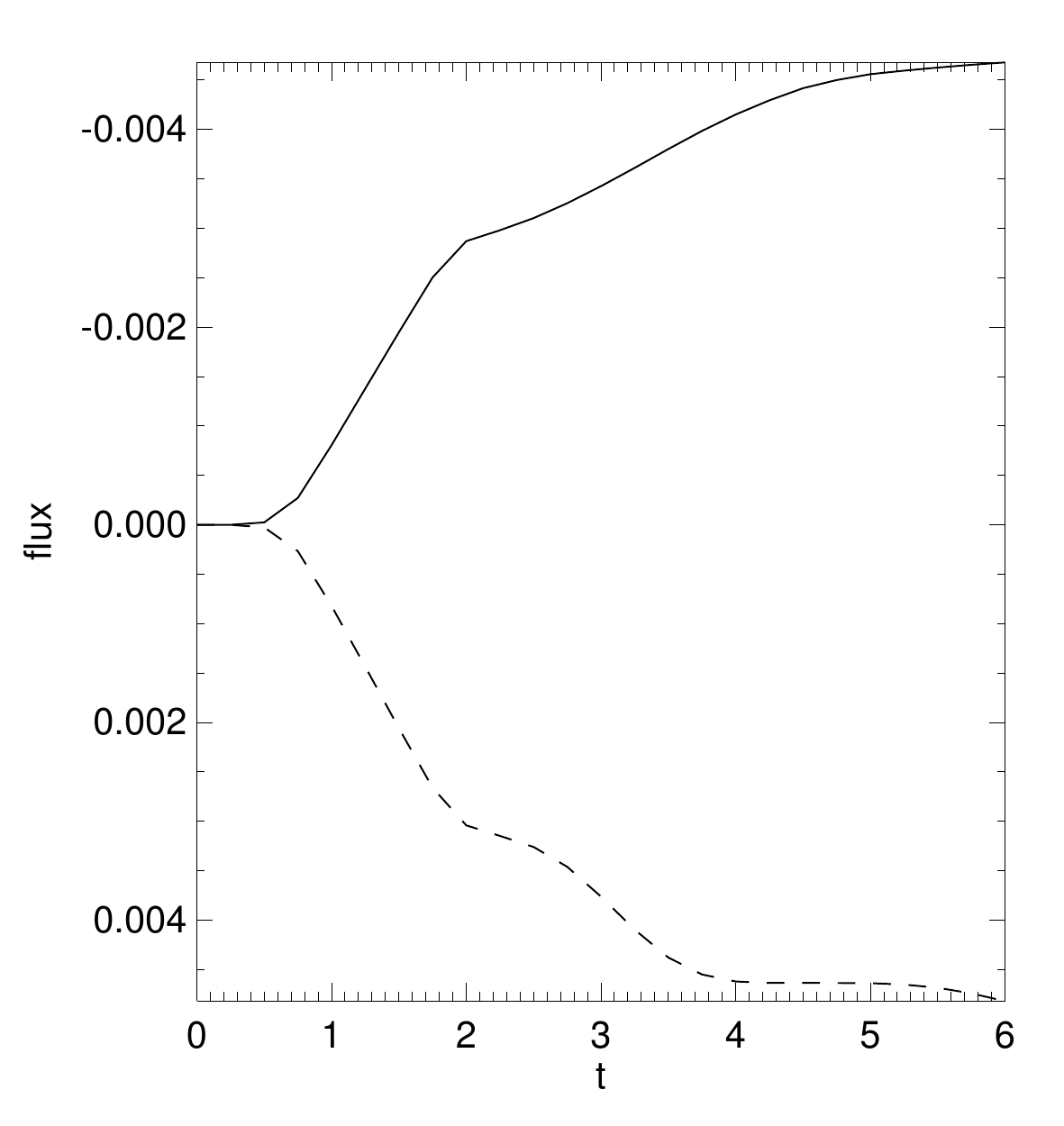}
\caption{(a) Map at $z=0$ showing connectivity change of field line footpoints for the MHD simulation. Field lines that lie inside the dome at $t=0$ but outside at $t=6$ emanate from the black region. Field lines originating in the white region are outside the dome at $t=0$ and inside at $t=6$. Field lines originating in the grey region remain either inside or outside the separatrix. (b) Cumulative flux entering (dashed line) and leaving (solid line) the dome over time.}
\label{fluxchange}
\end{figure}
The localised current concentration on the separatrix leads to a change in field line connectivities as seen in the simple model of the previous section. This can be observed by tracing field lines from the vicinity of the separatrix footprint during the simulation, and determining whether they lie inside or outside the dome at each time. Figure \ref{fluxchange}(a) shows the result of performing such a procedure using $10^6$ field lines. The black region of the resulting connectivity map shows the footpoint locations of field lines that are initially under the dome, but which end up outside the dome at $t=6$ following the reconnection process. The white region corresponds to field lines that begin outside the dome but end up inside it. We see clearly that the separatrix deforms to expel a region of flux on the flank closest to the null, while admitting flux in a narrow band around its remaining cirumference. We measure the flux entering and leaving the dome by counting field lines that enter and leave, weighting them by the local field strength at $z=0$ and the area element they correspond to with respect to neighbouring field lines. This is plotted in Figure \ref{fluxchange}(b). It is clear that the flux leaving the dome (solid line) is balanced by the flux entering the dome (dashed line) -- there are only very small discrepancies due to uncertainties and approximations used in this method. We see that between $t=1$ and $t=2$ there is a relatively sharp change in the flux, with this flux change gradually slowing down as the currents are dissipated.

\section{Conclusions}\label{concsec}
Three-dimensional magnetic null points are ubiquitous in the solar atmosphere, and in any generic mixed-polarity magnetic field. They are known to be susceptible to collapse leading to the generation of intense current layers at which, even with astrophysical plasma parameters, magnetic reconnection can occur. The simplest generic null configuration in a field above a conducting plane is that of a separatrix dome. Spine-fan reconnection at the null permits a transfer of flux from inside to outside the dome, and vice-versa. Importantly, there is no separator and yet still a transfer of flux between topologically distinct domains occurs. The null collapse and current sheet formation may be driven dynamically as in the MHD simulation described in Section \ref{numsec}, or may occur during a relaxation of the field \citep{pontin2012b}, and this flux transfer allows the magnetic field to lower its energy.

In the simple case without any flux emergence, the net flux through the photosphere beneath the dome is fixed (equal to the total flux associated with the parasitic polarity), so the reconnection process involves a balance between flux transferred into the dome on one region of the fan surface and flux transferred out on another region. The rate of this flux transfer is calculated by integrating the parallel electric field along the particular fan field line perpendicular to the plane of null collapse, this giving the maximum integrated parallel electric field. 

In the two models discussed above we studied field configurations in which the outer spine closes down to the photosphere. Thus, all flux close to the null was closed on the global scale (with some flux closing locally beneath the dome). However, one can also consider the situation where the outer spine is open to interplanetary space. In this case, spine-fan reconnection at the null involves the conversion of globally open flux to globally closed flux, and vice-versa. This has previously been termed ``interchange reconnection" \citep[e.g.][]{crooker2002,edmondson2010,masson2012}. This terminology is rather misleading, as it seems to suggest a direct one-to-one reconnection of field lines at the null. However, as discussed in Section \ref{recsec}, 3D reconnection occurs in a finite volume, not at a point, and there is no one-to-one correspondence between field line pairs before and after reconnection. Understanding the dynamics of the reconnection between open and closed magnetic flux requires that we understand the reconnection process occurring in the null point current sheet \citep[see also the discussion of][]{antiochos2007}.

As mentioned in Section \ref{recsec}, the magnetic connectivity change is by definition continuous during 3D reconnection. This is a direct consequence of the non-existence of a single flux velocity solving Equation (\ref{idealev}) in the presence of a localised non-ideal region. So the apparent flipping of magnetic field lines when traced from ideal co-moving footpoints through the non-ideal region is a natural feature of three-dimensional reconnection. When tracking field lines in this way, the change of footpoint mapping is discontinuous for field lines that pass exactly through the spine. Importantly, the flipping velocity ($\ww_{out}$ in the example of Section \ref{toysec}, displayed in Figures \ref{fliplines_sp} and \ref{flipmap}) is arbitrarily fast for field lines passing close to the spine. As a result we do not believe it to be physically meaningful to discuss the ``slip-running" speed of field lines in the presence of a null  \citep{masson2012,reid2012}. In particular, since the squashing factor $Q$ is infinite by definition at a null point, there will always be a finite region around the null of large $Q$ (a quasi-separatrix layer). However, the current will typically focus in a thin current sheet at the null, with reconnection occurring everywhere within this finite volume. This is the nature of 3D null point reconnection: as with all three-dimensional reconnection, it occurs in a finite volume and involves a continuous change of field line connectivity.

\acknowledgments
DP and ERP are each grateful to the Leverhulme Trust for financial support. Computations were carried out on the UKMHD consortium cluster funded by STFC and SRIF.

\bibliographystyle{apalike} 

\end{document}